\documentclass[twocolumn]{aastex631}

\usepackage{amsmath}

\usepackage{pbox, cellspace}
\usepackage{hyperref}
\renewcommand{\arraystretch}{2}

\begin{document}

\title{The First Radio-Bright Off-Nuclear TDE\,2024tvd Reveals the Fastest-Evolving Double-Peaked Radio Emission}

\correspondingauthor{Itai Sfaradi}
\email{itai.sfaradi@berkeley.edu}

\author[0000-0003-0466-3779]{Itai Sfaradi}
\affiliation{Department of Astronomy, University of California, Berkeley, CA 94720-3411, USA}
\affiliation{Berkeley Center for Multi-messenger Research on Astrophysical Transients and Outreach (Multi-RAPTOR), University of California, Berkeley, CA 94720-3411, USA}

\author[0000-0003-4768-7586]{Raffaella Margutti}
\affiliation{Department of Astronomy, University of California, Berkeley, CA 94720-3411, USA}
\affiliation{Department of Physics, University of California, 366 Physics North MC 7300, Berkeley, CA 94720, USA}
\affiliation{Berkeley Center for Multi-messenger Research on Astrophysical Transients and Outreach (Multi-RAPTOR), University of California, Berkeley, CA 94720-3411, USA}

\author[0000-0002-7706-5668]{Ryan Chornock}
\affiliation{Department of Astronomy, University of California, Berkeley, CA 94720-3411, USA}
\affiliation{Berkeley Center for Multi-messenger Research on Astrophysical Transients and Outreach (Multi-RAPTOR), University of California, Berkeley, CA 94720-3411, USA}

\author[0000-0002-8297-2473]{Kate D.~Alexander}
\affiliation{Department of Astronomy and Steward Observatory, University of Arizona, 933 North Cherry Avenue, Tucson, AZ 85721-0065, USA}

\author[0000-0002-4670-7509]{Brian D.~Metzger}
\affiliation{Department of Physics and Columbia Astrophysics Laboratory, Columbia University, New York, NY 10027, USA}
\affiliation{Center for Computational Astrophysics, Flatiron Institute, 162 5th Ave, New York, NY 10010, USA}

\author[0000-0001-7833-1043]{Paz Beniamini}
\affiliation{Department of Natural Sciences, The Open University of Israel, P.O Box 808, Ra'anana 4353701, Israel}
\affiliation{Astrophysics Research Center of the Open University (ARCO), The Open University of Israel, P.O Box 808, Ra'anana 4353701, Israel}
\affiliation{Department of Physics, The George Washington University, 725 21st Street NW, Washington, DC 20052, USA}

\author[0000-0002-5565-4824]{Rodolfo Barniol Duran}
\affiliation{Department of Physics and Astronomy, California State University, Sacramento, 6000 J Street, Sacramento CA 95819-6041 USA}

\author[0000-0001-6747-8509]{Yuhan Yao}
\affiliation{Department of Astronomy, University of California, Berkeley, CA 94720-3411, USA}
\affiliation{Miller Institute for Basic Research in Science, 206B Stanley Hall, Berkeley, CA 94720, USA}
\affiliation{Berkeley Center for Multi-messenger Research on Astrophysical Transients and Outreach (Multi-RAPTOR), University of California, Berkeley, CA 94720-3411, USA}

\author[0000-0002-5936-1156]{Assaf Horesh}
\affiliation{Racah Institute of Physics, The Hebrew University of Jerusalem, Jerusalem 91904, Israel}

\author[0000-0002-0161-7243]{Wael Farah}
\affiliation{Department of Astronomy, University of California, Berkeley, CA 94720-3411, USA}
\affiliation{SETI Institute, 339 Bernardo Ave, Suite 200 Mountain View, CA 94043, USA}
\affiliation{Berkeley SETI Research Centre, University of California, Berkeley, CA 94720, USA}

\author[0000-0002-9392-9681]{Edo Berger}
\affiliation{Center for Astrophysics \textbar{} Harvard \& Smithsonian, 60 Garden Street, Cambridge, MA 02138-1516, USA}

\author[0000-0002-8070-5400]{Nayana A. J.}
\affiliation{Department of Astronomy, University of California, Berkeley, CA 94720-3411, USA}
\affiliation{Berkeley Center for Multi-messenger Research on Astrophysical Transients and Outreach (Multi-RAPTOR), University of California, Berkeley, CA 94720-3411, USA}

\author[0000-0001-7007-6295]{Yvette Cendes}
\affiliation{Department of Physics, University of Oregon, 1371 E 13th Ave, Eugene OR 97403, USA}
\affiliation{Institute for Fundamental Science, University of Oregon, 1371 E 13th Ave, Eugene OR 97403, USA}

\author[0000-0003-0307-9984]{Tarraneh Eftekhari}
\affiliation{Center for Interdisciplinary Exploration and Research in Astronomy (CIERA), Northwestern University, 1800 Sherman Avenue, Evanston, IL 60201, USA}

\author{Rob Fender}
\affiliation{Astrophysics, Department of Physics, University of Oxford, Keble Road, Oxford, OX1 3RH, UK}

\author[0000-0003-4537-3575]{Noah Franz}
\affiliation{Department of Astronomy and Steward Observatory, University of Arizona, 933 North Cherry Avenue, Tucson, AZ 85721-0065, USA}

\author[0000-0003-3189-9998]{Dave~A.~Green}
\affiliation{Astrophysics Group, Cavendish Laboratory, J. J. Thomson Avenue, Cambridge CB3 0US, UK}

\author[0000-0002-5698-8703]{Erica Hammerstein}
\affiliation{Department of Astronomy, University of California, Berkeley, CA 94720-3411, USA}
\affiliation{Berkeley Center for Multi-messenger Research on Astrophysical Transients and Outreach (Multi-RAPTOR), University of California, Berkeley, CA 94720-3411, USA}

\author[0000-0002-1568-7461]{Wenbin Lu}
\affiliation{Department of Astronomy, University of California, Berkeley, CA 94720-3411, USA}
\affiliation{Theoretical Astrophysics Center, University of California, Berkeley, CA 94720-3411, USA}
\affiliation{Berkeley Center for Multi-messenger Research on Astrophysical Transients and Outreach (Multi-RAPTOR), University of California, Berkeley, CA 94720-3411, USA}

\author[0000-0000-0000-0000]{Eli Wiston}
\affiliation{Department of Astronomy, University of California, Berkeley, CA 94720-3411, USA}
\affiliation{Berkeley Center for Multi-messenger Research on Astrophysical Transients and Outreach (Multi-RAPTOR), University of California, Berkeley, CA 94720-3411, USA}

\author{Yirmi Bernstein}
\affiliation{Racah Institute of Physics, The Hebrew University of Jerusalem, Jerusalem 91904, Israel}

\author[0000-0002-7735-5796]{Joe Bright}
\affiliation{Astrophysics, Department of Physics, University of Oxford, Keble Road, Oxford, OX1 3RH, UK}

\author[0000-0003-0528-202X]{Collin~T.~Christy}
\affiliation{Department of Astronomy and Steward Observatory, University of Arizona, 933 North Cherry Avenue, Tucson, AZ 85721-0065, USA}

\author[0000-0001-5576-2254]{Luigi F. Cruz}
\affiliation{SETI Institute, 339 Bernardo Ave, Suite 200 Mountain View, CA 94043, USA}

\author[0000-0003-3197-2294]{David R DeBoer}
\affiliation{Radio Astronomy Laboratory, University of California, Berkeley, CA, 94720 USA}
\affiliation{Sub-department of Astrophysics, University of Oxford, Oxford, OX1-3RH, UK}

\author[0000-0001-7946-1034]{Walter W. Golay}
\affiliation{Center for Astrophysics \textbar{} Harvard \& Smithsonian, 60 Garden Street, Cambridge, MA 02138-1516, USA}

\author[00000-0003-3441-8299]{Adelle J. Goodwin}
\affiliation{International Centre for Radio Astronomy Research -- Curtin University, GPO Box U1987, Perth, WA 6845, Australia}

\author[0000-0003-0685-3621]{Mark Gurwell}
\affiliation{Center for Astrophysics \textbar{} Harvard \& Smithsonian, 60 Garden Street, Cambridge, MA 02138-1516, USA}

\author[0000-0002-3490-146X]{Garrett~K.~Keating}
\affiliation{Center for Astrophysics \textbar{} Harvard \& Smithsonian, 60 Garden Street, Cambridge, MA 02138-1516, USA}

\author[0000-0003-1792-2338]{Tanmoy Laskar}
\affiliation{Department of Physics \& Astronomy, University of Utah, Salt Lake City, UT 84112, USA}

\author[0000-0003-3124-2814]{James C. A. Miller-Jones}
\affiliation{International Centre for Radio Astronomy Research -- Curtin University, GPO Box U1987, Perth, WA 6845, Australia}

\author[0000-0002-3430-7671]{Alexander~W.~Pollak}
\affiliation{SETI Institute, 339 Bernardo Ave, Suite 200 Mountain View, CA 94043, USA}

\author[0000-0002-1407-7944]{Ramprasad~Rao}
\affiliation{Center for Astrophysics \textbar{} Harvard \& Smithsonian, 60 Garden Street, Cambridge, MA 02138-1516, USA}

\author{Andrew Siemion}
\affiliation{Astrophysics, Department of Physics, University of Oxford, Keble Road, Oxford, OX1 3RH, UK}
\affiliation{Breakthrough Listen, Astrophysics, Department of Physics, The University of Oxford, Keble Road, Oxford OX1 3RH, UK}
\affiliation{SETI Institute, 339 Bernardo Ave, Suite 200 Mountain View, CA 94043, USA}
\affiliation{Berkeley SETI Research Centre, University of California, Berkeley, CA 94720, USA}
\affiliation{Department of Physics and Astronomy, University of Manchester, UK}
\affiliation{University of Malta, Institute of Space Sciences and Astronomy, Msida, MSD2080, Malta}

\author[0000-0001-7057-4999]{Sofia Z. Sheikh}
\affiliation{SETI Institute, 339 Bernardo Ave, Suite 200 Mountain View, CA 94043, USA}
\affiliation{Berkeley SETI Research Centre, University of California, Berkeley, CA 94720, USA}

\author[0009-0005-5622-3611]{Nadav Shoval}
\affiliation{Racah Institute of Physics, The Hebrew University of Jerusalem, Jerusalem 91904, Israel}

\author[0000-0002-3859-8074]{Sjoert van Velzen}
\affiliation{Leiden Observatory, Leiden University, Postbus 9513, 2300 RA, Leiden, The Netherlands}

\begin{abstract}

We present the first multi-epoch broadband radio and millimeter monitoring of an off-nuclear TDE using the VLA, ALMA, ATA, AMI-LA, and the SMA. The off-nuclear TDE\,2024tvd exhibits double-peaked radio light curves and the fastest evolving radio emission observed from a TDE to date. With respect to the optical discovery date, the first radio flare rises faster than $F_{\rm \nu} \sim t^{9}$ at $\Delta t = 88-131$ days, and then decays as fast as $F_{\rm \nu} \sim t^{-6}$. The emergence of a second radio flare is observed at $\Delta t \approx 194$ days with an initial fast rise of $F_{\rm \nu} \sim t^{18}$, and an optically thin decline of $F_{\rm \nu} \sim t ^{-12}$. We interpret these observations in the context of a self-absorbed and free--free absorbed synchrotron spectrum, while accounting for both synchrotron and inverse-Compton cooling. We find that a single prompt outflow cannot easily explain these observations and it is likely that either there is only one outflow that was launched at $\Delta t \sim 80$ days, or two distinct outflows, with the second launched at $\Delta t \sim 170-190$ days. The nature of these outflows, whether sub-, mildly-, or ultra-relativistic, is still unclear, and we explore these different scenarios. Finally, we find a temporal coincidence between the launch time of the first radio-emitting outflow and the onset of a power-law component in the X-ray spectrum, attributed to inverse-Compton scattering of thermal photons.

\end{abstract}

\keywords{Tidal disruption (1696) --- Supermassive black holes (1663) --- Radio astronomy (1338) ---  Time domain astronomy (2109)} 

\section{Introduction}
\label{sec:intro}

Tidal disruption events (TDEs) occur when a star is torn apart by the extreme tidal forces of a massive black hole \citep[MBH;][]{hills_1975, Rees_1988}. These cataclysmic events produce radiation across the electromagnetic spectrum, providing valuable information on the demographics of MBHs, the resulting accretion processes, the fast outflows generated during stellar disruption and disk formation, and the various physical mechanisms governing TDEs \citep{alexander_2020, van_velzen_2021, Hammerstein_2023, Yao_2023,guolo_2024}. Most TDEs are observed in galactic nuclei, with a small offset from the host nucleus serving as a classification criterion in many cases \citep{Hammerstein_2023}. However, a distinct subset known as off-nuclear TDEs occurs outside the central regions of galaxies and can be associated with wandering \citep{ricarte_2021a, ricarte_2021b} or recoiling \citep{Stone_2011} black holes, with only a handful of off-nuclear TDE-candidates discovered so far \citep{Lin_2018, Lin_2020, Jin_2025, Yao_2025, goodwin_erosite, grotova_erosita, guolo_hlx_2025}. Studying these events presents a unique opportunity to probe MBH populations beyond galactic nuclei, investigate the potential presence of intermediate-mass black holes, and explore the dynamics of MBHs before and after coalescence in non-traditional environments.

The interaction between the fast outflows from TDEs (e.g., unbound tidal debris stream, winds from the accretion disk, and relativistic jets) and the surrounding medium can generate shocks, producing non-thermal synchrotron emission (e.g., \citealt{alexander_2020}). Since this non-thermal emission typically peaks at radio wavelengths, radio observations play a key role in studying the interaction region. Early-time radio emission, detected within the first few weeks to months after optical discovery, has been observed in approximately $30\%$ of optically detected TDEs \citep{alexander_2020}, revealing a wide range of outflow properties. For example, the kinetic energy associated with these outflows spans several orders of magnitude, from $\sim 10^{48} - 10^{50} \, \rm erg$ for sub-relativistic and mildly relativistic outflows (e.g., ASASSN-14li; \citealt{Alexander_2016, Krolik_2016}, and AT\,2019dsg; \citealt{Stein_2021, Cendes_2022}) to $\sim 10^{51} - 10^{53} \, \rm erg$ for relativistic jets (e.g., Swift J1644+57; \citealt{Zauderer_2011, berger_2012, Eftekhari_2018, Beniamini_2023_J1644}; AT\,2022cmc; \citealt{Andreoni_2022, Rhodes_2023}; and the possible off-axis relativistic jet from AT\,2018hyz; \citealt{Matsumoto_2023, Sfaradi_2024,Cendes_2025}).

Sub-relativistic outflows, with typical velocities of up to a few $0.1c$, are typically associated with the unbound stellar debris stream, accretion-driven winds, or stream--stream collisions (e.g., ASASSN-14li; \citealt{Alexander_2016}, the first flare from AT\,2020vwl; \citealt{Goodwin_2023}, and potentially AT\,2019dsg; \citealt{Cendes_2022}; see also \citealt{Stein_2021} for an alternative explanation). If these outflows originate from the unbound stellar debris or accretion disk winds, the resulting radio emission provides a rare opportunity to probe the circum-nuclear medium (CNM) on scales of $10^{15} - 10^{17} \, \rm cm$. Relativistic jets probe the CNM, or the interstellar medium (ISM), at much larger distances, typically $10^{17} - 10^{19} \, \rm cm$. On the other hand, radio emission from stream-stream collisions can shed light on the density structure of the unbound tidal debris stream.

Approximately $40\%$ of optically discovered TDEs exhibit late-time radio brightening months to years after stellar disruption \citep{Cendes_2024,Alexander_2025}. Additionally, some TDEs show multiple flares in their radio light curves on different timescales (e.g., ASASSN-15oi; \citealt{Horesh_2021a, Hajela_2025}, AT\,2019azh; \citealt{Sfaradi_2022}, and AT\,2020vwl; \citealt{Goodwin_2023}). Several mechanisms have been proposed to explain these late-time flares, and they are broadly divided into (a) those that invoke an outflow that is launched around the time of optical discovery and is either interacting with a complex density structure \citep{Horesh_2021b, Matsumoto_2024} or is initially pointing away from our line of sight \citep{Matsumoto_2023, Sfaradi_2024, Christy_2024, Cendes_2025}, and (b) those that require a delayed launch of an outflow (potentially accretion-driven outflow; \citealt{Giannios_2011,  Horesh_2021a, Sfaradi_2022, Cendes_2022, Piro_2025, goodwin_2025}).

TDE-driven fast outflows also provide a natural laboratory for studying the microphysics of both non-relativistic and relativistic shock waves. Electrons at the shock front are accelerated to relativistic velocities, typically forming a power-law energy distribution. Observations of the optically-thin regime of the synchrotron spectrum offer a unique probe of this power-law index, which encodes information about the process of particle acceleration (e.g., \citealt{caprioli_2023}). A fraction of the post-shock energy is divided between the electrons and the magnetic field, often assumed to be in equipartition for the lack of better observational guidance \citep{chevalier_1998, BDNP13}. However, when measurements of the synchrotron cooling spectral break are available, deviations from equipartition are often observed (e.g., AT\,2019dsg; \citealt{cendes_2021}; AT\,2018hyz; \citealt{Cendes_2022}; ASASSN-19bt; \citealt{Christy_2024}).

So far, radio observations have been reported for only three off-nuclear intermediate mass black hole (IMBH) TDEs and TDE-candidates. The IMBH-TDE EP240222a was not detected with a $5\sigma$ upper limit of $\nu L_{\rm \nu} \lesssim 10^{37} \, \rm erg \, s^{-1}$ \citep{Jin_2025}. The other two IMBH TDE candidates, HLX-1 and eRASSt J142140-295321, revealed low level of radio luminosity of $\nu L_{\nu} \simeq 5 \times 10^{36}$ and $3 \times 10^{37} \, \rm erg \, s^{-1}$, respectively \citep{webb_2012, goodwin_erosite}. However, the classification of these two transients as IMBH-TDEs remains uncertain. This paper discusses AT\,2024tvd, the first off-nuclear TDE selected from optical sky surveys, which is also the first bonafide off-nuclear TDE with bright radio emission.

AT\,2024tvd was discovered in the Zwicky Transient Facility \citep[ZTF;][]{bellm_2019, graham_2019}) $g$-band at a magnitude of $19.68$ on August 25, 2024 \citep{24tvd_discovery_tns} with the 48-inch Samuel Oschin Schmidt telescope at Palomar Observatory (P48). It was then classified as a TDE based on the broad H and \ion{He}{2} in the spectrum, its central location in its host galaxy, and the long-lasting UV emission \citep{Faris_24tvd}. The off-nuclear position of AT\,2024tvd was first reported by \cite{24tvd_yao_tns}, and a MBH mass of $10^{5} -10^{7} \, M_{\odot}$ at an offset of $0.8$ kpc from the host nucleus was then estimated by \cite{Yao_2025}. The origin of the off-nuclear position of this TDE is likely to be a recoiling MBH from a triple MBH interaction or a MBH in the least massive galaxy of a minor galaxy merger. The reported redshift of AT\,2024tvd is $z=0.04494$, and we adopt a luminosity distance of $200$ Mpc based on a standard $\Lambda$CDM cosmology with $\Omega_{\rm M} = 0.3$, $\Omega_{\Lambda} = 0.7$, and a Hubble constant $H_0 = 70 \, \rm km \, s^{-1} \, Mpc^{-1}$. All times and dates are given in UT.

In this paper, we present comprehensive radio monitoring of AT\,2024tvd (\S\ref{sec:radio_observations}), and we find it to be the first radio bright off-nuclear TDE, and the TDE with the fastest evolution of the radio emission observed to date (\S\ref{sec: AT2024tvd}). We model the radio emission from AT\,2024tvd in the context of a self-absorbed synchrotron model, and account (for the first time for a TDE in the radio) for both free--free absorption by the material in-front of the radio emitting shock, and inverse-Compton cooling by the thermal optical/UV photons (see \S\ref{sec:analysis}). We discuss the astrophysical implications of our analysis in \S\ref{sec: discussion}, and present our conclusions in \S\ref{sec:conclusions}.

\section{Radio observations}
\label{sec:radio_observations}

Following the optical classification of AT\,2024tvd as a TDE \citep{Faris_24tvd}, and the analysis of its off-nuclear position (first reported by \cite{24tvd_yao_tns}; see also \S\ref{sec: AT2024tvd}), we triggered the Karl G. Jansky Very Large Array (VLA) on 2024 November 21, which corresponds to $\Delta t= 88$ days ($\Delta t$ is the time since optical discovery in the frame of the observer), under our approved program to follow-up TDEs (VLA 24A-386; PI: Alexander). This X-band observation ($\nu = 10$ GHz) resulted in a non-detection at the coordinates of the TDE. Instead, a point source emission at the coordinates of the host's nucleus was detected (see left panel of Fig.~\ref{fig: field_images}). Broadband follow-up observations in S--Ku bands (centered around 3--15 GHz) of the field of AT\,2024tvd, conducted on 2025 January 3, corresponding to $\Delta t = 131$ days, showed a point source at the position of the TDE, and, a point source at the position of the host galaxy nucleus in $3$, $6$, and $10$ GHz (we first reported the $10$ GHz detection from this observation in \citealt{24tvd_sfaradi_tns} and present a $6$ GHz image in the right panel of Fig.~\ref{fig: field_images}). Following this detection, we initiated a broadband, multi-epoch campaign, using the VLA, Atacama Large Millimeter/submillimeter Array (ALMA), the Arcminute Micro-Kelvin Imager - Large Array (AMI-LA; \citealt{Zwart_2008, Hickish_2018}), the Allen Telescope Array (ATA; Farah et al. in prep, Pollak et al. in prep), and the Submillimeter Array (SMA). We emphasize here that our analysis takes into account the underlying emission from the host galaxy nucleus for observations that do not resolve the transient and the host (see detailed discussions in \S\ref{subsec:VLA}, \S\ref{subsec:ATA}, and Appendix \ref{sec: host_emission}).

\subsection{Karl G. Jansky Very Large Array}
\label{subsec:VLA}

We observed the field of AT\,2024tvd with the VLA under our dedicated TDE program (VLA 24A-386; PI: Alexander), and a DDT (VLA 25A-483; PI: Sfaradi). In all of our observations we used J1735+3616 as a phase calibrator. 3C147 and 3C48 were used as a band-pass and absolute flux calibrators. Our first four observations were conducted while the VLA was in its most extended A-configuration. The fifth observation with the VLA was conducted with the transitional A$\rightarrow$D configuration. All of the following observations were made while the VLA was in the more compact C- and D-configurations. We used the Common Astronomy Software Applications (CASA; \citealt{CASA}) packages and the VLA calibration pipeline to flag and calibrate the data. Additional manual flagging was applied when needed. 

\begin{figure*}[ht]
\includegraphics[width=\linewidth]{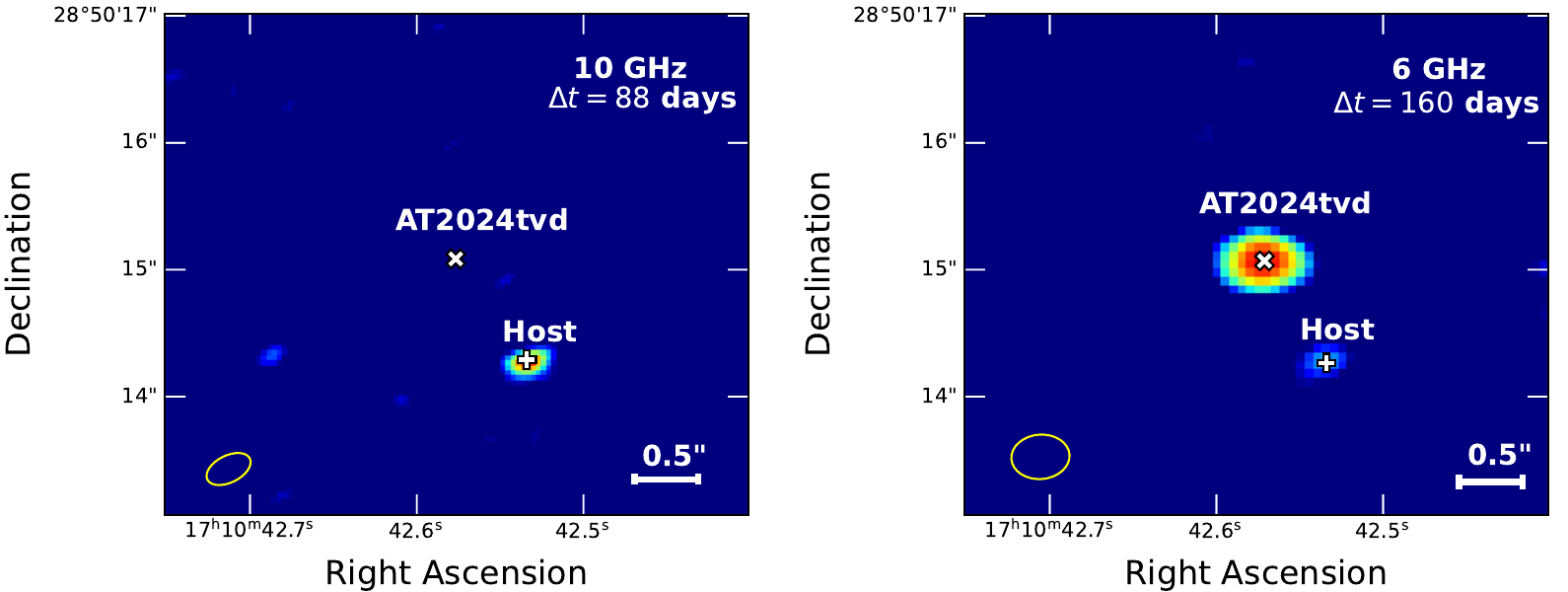}
\caption{Radio images of the field of AT\,2024tvd. \emph{Left panel}: the first VLA observation in X-band at $\Delta t = 88$ days, resulting in a null-detection of the TDE, and a detection of the nucleus of the host. \emph{Right panel}: the VLA C-band image at $\Delta t = 160$ days, showing both the emission from the TDE and from the host galaxy. The positions of the TDE and of the nucleus obtained from the HST image \citep{Yao_2025} are marked with an 'X' and a '+' sign, respectively. The size of the VLA clean beams is shown with a yellow ellipse at the bottom left of each VLA image. The VLA images are scaled differently in order to show the full range of flux densities from the TDE and from the host galaxy nucleus.
\label{fig: field_images}}
\end{figure*}

We used the CASA task \texttt{TCLEAN} to produce clean images of the field of AT\,2024tvd. Our S- and C-band images taken in A-configuration showed both the emission from the host nucleus and the TDE (see Fig.~\ref{fig: field_images}). When producing images while the VLA was in A-configuration we measure the flux density of both the transient and the host, if present, by using the CASA task \texttt{IMFIT}, and the task \texttt{IMSTAT} to calculate the image rms. We then estimate the error of the flux density to be a quadratic sum of the error produced by the CASA task \texttt{IMFIT}, and a $10\%$ calibration error \citep{Weiler_1986}. For observations taken in the VLA C- or D-configurations we measure only the flux density from the point source coincident with the TDE. Since we have measurements of the flux density from the nucleus, and since the flux density measured from these compact configurations images is contaminated by the host emission, when modeling the emission from the TDE we subtracted a power-law fit to the host nucleus emission as discussed in Appendix \S\ref{sec: host_emission}.

Our observation from 2025 February 14, at $\Delta t = 173$ days, was conducted in an hybrid array, $\rm A\rightarrow D$. We then used only the long baselines by setting the \texttt{TCLEAN} parameter $\texttt{uvrange} > 15 \, \rm k \lambda$ to minimize the level of host contamination. While this procedure results in an elongated beam, it allows us to separate the host emission from the TDE emission. For this observation, we measure the flux density as the flux at the position of the optical transient (and not fitting a point source with \texttt{IMFIT}), and estimate the uncertainty to be a quadratic sum of the image rms calculated by the CASA task \texttt{IMSTAT}, and a more conservative $15\%$ calibration error (to account for possible systematic errors when using only long baselines).

We provide the flux density measurements in Table~\ref{tab: radio_observations}, and different images of the field of AT\,2024tvd in Fig.~\ref{fig: field_images}.

\begin{deluxetable*}{cccccc}[ht]
\tablecaption{Summary of the radio flux measurements.}
\tablehead{
\colhead{Observation date} & \colhead{$\Delta t$} & \colhead{$\nu$} & \colhead{$F_{\rm \nu}$} &
\colhead{Image RMS} &
\colhead{Telescope} \\
\colhead{[DD-MM-YYYY]} & \colhead{$\rm \left[ Days \right]$} & \colhead{$\rm \left[ GHz \right]$} & \colhead{$\rm \left[ mJy \right]$} & \colhead{$\rm \left[ mJy \right]$} & \colhead{}}
\startdata
21-11-2024 & $88$ & $10$ & $\leq 0.0165$ & $0.0055$ & VLA:A \\
03-01-2025 & $131$ & $3$ & $0.06 \pm 0.02$ & $0.013$ & VLA:A \\
03-01-2025 & $131$ & $6$ & $0.27 \pm 0.03$ & $0.008$ & VLA:A \\
03-01-2025 & $131$ & $10$ & $0.60 \pm 0.06$ & $0.011$ & VLA:A \\
03-01-2025 & $131$ & $13$ & $0.90 \pm 0.10$ & $0.018$ & VLA:A \\
03-01-2025 & $131$ & $15$ & $1.01 \pm 0.10$ & $0.015$ & VLA:A \\
03-01-2025 & $131$ & $17$ & $1.10 \pm 0.11$ & $0.014$ & VLA:A \\
11-01-2025 & $139$ & $97.5$ & $0.22 \pm 0.03$ & $0.027$ & ALMA \\
\enddata
\tablecomments{Summary of the radio flux density measurements for the off-nuclear TDE\,2024tvd. $\Delta t$ is the time since optical discovery (in the observer frame), $\nu$ is the observed central frequency, $F_{\rm \nu}$ is the flux density (and $1 \sigma$ uncertainty), the image rms is also reported. In the ``Telescope'' column we specify different VLA configurations. Note that when analyzing VLA observations obtained in C- and D-configuration we removed a power-law fit to the observed host nucleus emission (see Appendix \S\ref{sec: host_emission}). In addition, we cross-calibrated the ATA fluxes at $\Delta t = 249$ days with the VLA, and removed the excess emission from the ATA observations due to the contamination from the larger beam of the ATA. A full version of this table is accessible online in a machine readable table format.
\label{tab: radio_observations}}
\end{deluxetable*}

\subsection{Atacama Large Millimeter/submillimeter Array}
\label{subsec:ALMA}

Following our first detection with the VLA we obtained mm-band observations with ALMA in four epochs under two DDT programs (2024.A.00024.T; 2024.A.00034.T; PI: Sfaradi). We used ALMA Bands 3 and 5 with the central frequencies of $97.5$ and $203$ GHz, respectively, J1550+0527 as an absolute flux and bandpass calibrator, and J1753+2848 as a phase calibrator. We used the standard National Radio Astronomy Observatory (NRAO) calibrated images, and fitted the point source with CASA task \texttt{IMFIT}, and used CASA task \texttt{IMSTAT} to obtain the image rms. We provide the flux density measurements in Table~\ref{tab: radio_observations}.

\subsection{Arcminute Micro-Kelvin Imaging - Large Array}
\label{subsec:AMI}

AMI-LA is a radio interferometer that consists of eight antennas, each $12.8 \, \rm m$ in diameter, with a maximum separation of $110 \, \rm m$. It observes with a $5$ GHz bandwidth around a central frequency of $15.5$ GHz. We observed the field of AT\,2024tvd with AMI-LA with high cadence starting 2025 January 12, at $\Delta t = 140$ days, and reported the first detection in \cite{horesh_24tvd}. We reduced, flagged, and calibrated our observations using \texttt{reduce\_dc}, a
customized AMI-LA data reduction software package \citep{perrott_2013}. Images of the field of AT\,2024tvd were produced using CASA task \texttt{TCLEAN} in an interactive mode. We fit the source at the optical position of the TDE with the CASA task \texttt{IMFIT} and calculated the image rms with the CASA task \texttt{IMSTAT}. We estimate the error of the peak flux density to be a quadratic sum of the error produced by the CASA task \texttt{IMFIT}, and a $10\%$ calibration error. We provide the flux density measurements in Table~\ref{tab: radio_observations}. 

\subsection{Allen Telescope Array}
\label{subsec:ATA}

We observed the field of AT\,2024tvd with the ATA under our program (P062; PI: Sfaradi) starting on 2025 March 20, which corresponds to $\Delta t = 207$ days. The ATA is a radio interferometer that comprises 42 dishes, each with a diameter of 6.1 m and can utilize up to four independent frequency tunings in the range of $1-10$ GHz, each frequency tuning with a $\sim 700$ MHz width \citep{Bright_2023}. Our observations were centered around $1.5$, $3$, $5$, and $8$ GHz using 3C286 to calibrate the absolute flux scale and band-pass response, and 1735+362 to calibrate the time-dependent complex gains. We used a customized pipeline (\url{https://github.com/joesbright/ATARI/}) utilizing CASA to reduce the observations, and CASA task \texttt{TCLEAN} for imaging.

The flux density measured by the ATA is contaminated by the background due to the large beam of the ATA. Therefore, we cross-calibrated all the flux density measurements obtained with the ATA by removing the excess emission seen in the ATA observation on 2025 May 1 (at $\Delta t = 249$ days), compared to the flux density measured by the VLA on 2025, April 29 (at $\Delta t = 247$ days). We estimate the error of the peak flux density to be a quadratic sum of the error produced by CASA task \texttt{IMFIT}, and $15\%$ calibration error. We provide the flux density measurements in Table~\ref{tab: radio_observations}. 

\subsection{Submillimeter Array}
\label{subsec: SMA}

Beginning 2025 April 21 ($\Delta t = 239$ days), several observations of AT\,2024tvd were conducted with the SMA, as part of the Pursuit of Extragalactic Transients with the SMA (POETS) Large-Scale SMA program (2022B-S046; PI: E. Berger). SMA observations were conducted with the array in compact configuration ($\theta_{\rm b} = 3.5" \times 3.2"$), tuned to a local oscillator (LO) frequency of $225.5$ GHz, providing spectral coverage between $209.5-221.5$ and $229.5-241.5$ GHz. For all observations, 3C279 was observed as a bandpass calibrator, Vesta was observed as the flux calibrator, and 1642+398 and 1658+347 were observed as gain calibrators, with a 12 minute cycle time cadence.

Analysis of the data was performed using the SMA COMPASS pipeline (G. K. Keating et al, in prep), which derives bandpass and gains calibration tables, flags outlier visibilities, as well baselines where limited or no coherence is seen on calibration targets. Flux calibration uses the Butler–JPL–Horizons 2012 \citep{butler_2012} model for Vesta. The data were imaged, and de-convolution was performed via the CLEAN algorithm \citep{Hogbom1974}. We provide the flux density measurements in Table~\ref{tab: radio_observations}. 

\section{Radio emission from AT 2024tvd: broad considerations}
\label{sec: AT2024tvd}

Next, we present the entire set of radio observations we have obtained at $\Delta t \leq 284$ days. We begin by examining the overall evolution of the radio emission from AT\,2024tvd over $\sim 200$ days, with observations spanning from the cm to the mm bands. We discuss the emission from the host galaxy nucleus in Appendix \ref{sec: host_emission}. In Fig.~\ref{fig: radio_comparison} we plot the $10$ GHz light-curve of AT\,2024tvd compared to other known radio-bright TDEs. Fig.~\ref{fig: light_curves} presents light curves in different cm- and mm-bands together with temporal power-law indices with respect to the optical discovery date. In Fig.~\ref{fig: radio_observations} we present the different spectral energy distributions (SEDs) with the best-fitting broken power-law functions (see detailed discussion in \S\ref{subsec: individual_analysis}). Overall, as seen from these plots, this TDE exhibits two episodes of fast-evolving radio emission, with distinct broadband evolution in each flare.

\begin{figure}[ht]
\includegraphics[width=\linewidth]{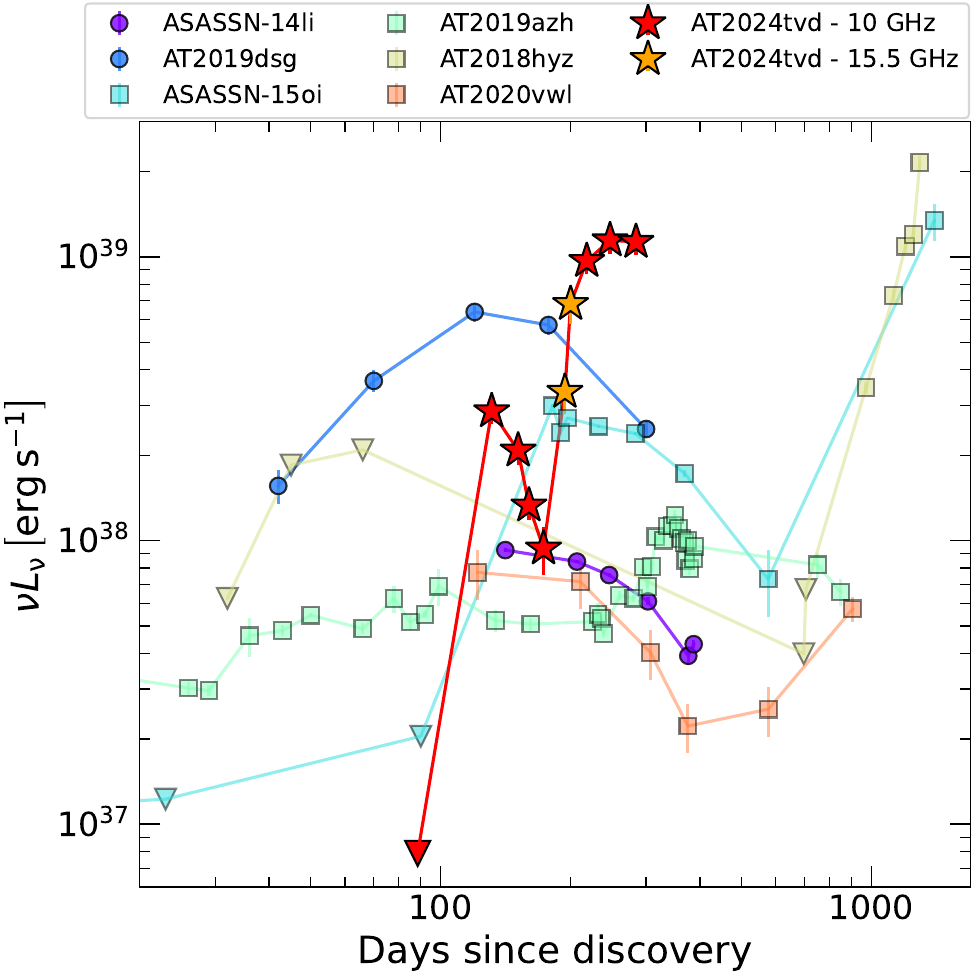}
\caption{Radio luminosity evolution of a selection of known TDEs (ASASSN-14li; \citealt{Alexander_2016}, ASASSN-15oi; \citealt{Horesh_2021a}, AT\,2019dsg; \citealt{Stein_2021}, AT\,2018hyz; \citealt{Cendes_2022}, AT\,2019azh; \citealt{Goodwin_2022,Sfaradi_2022}, and AT\,2020vwl; \citealt{Goodwin_2023}), and AT\,2024tvd (this work; $10$ GHz marked with red stars; $15.5$ GHz marked with orange stars). Observations of the same TDE are connected with lines, and $3\sigma$ upper limits are plotted with triangles. The radio emission from the off-nuclear TDE\,2024tvd reveals an extremely fast evolution compared to other TDEs, with a fast rise between $\Delta t = 88$ and $131$ days, followed by a fast decline up to $\Delta t = 173$ days. Then, at $\Delta t = 194$ days, and about $\sim 20$ days since the last observation of the first radio flare, a second fast-rising radio flare emerges.
\label{fig: radio_comparison}}
\end{figure}

\begin{figure}[ht]
\includegraphics[width=\linewidth]{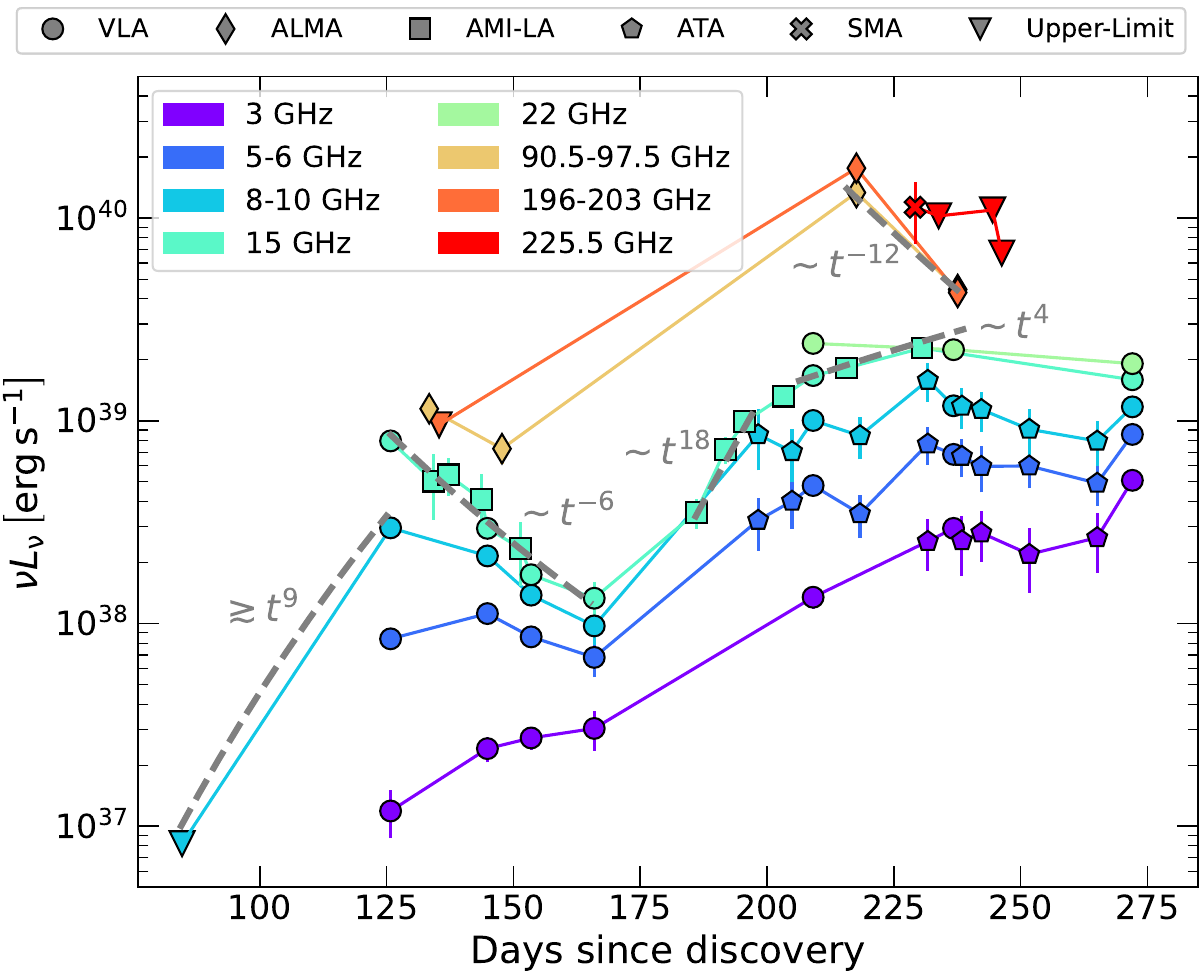}
\caption{Radio light curves of the off-nuclear TDE\,2024tvd at different frequencies. We plot temporal power-laws for different segments of these light curves (with the reference initial time being the optical discovery), showing the fast rise and decay in both flares.
\label{fig: light_curves}}
\end{figure}

\begin{figure*}[ht]
\includegraphics[width=\linewidth]{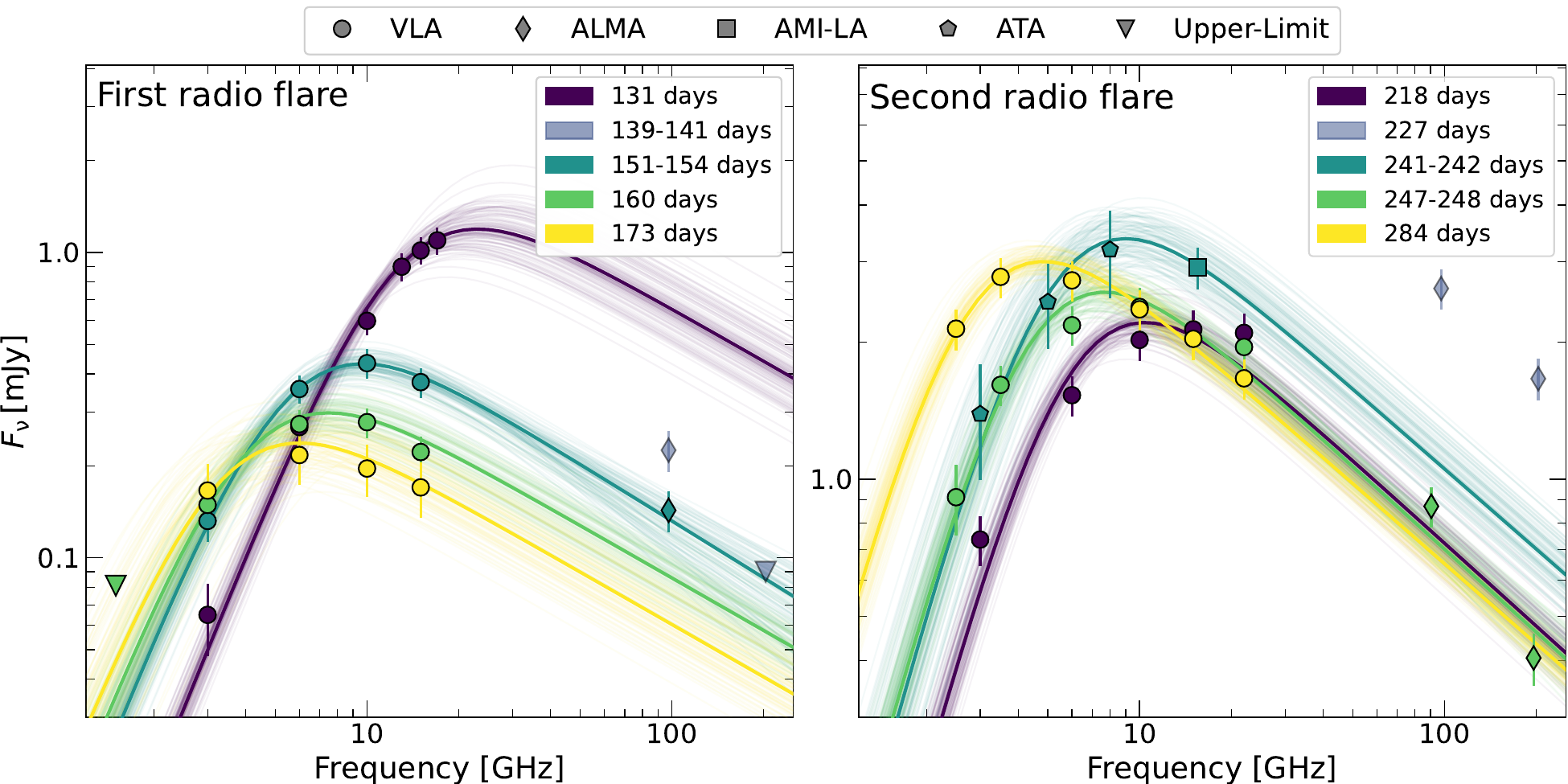}
\caption{Broadband radio SEDs of the off-nuclear TDE\,2024tvd. The left panel shows the radio SEDs of the first radio flare, and the right panel is for the second radio flare. Detections obtained with different radio telescopes are marked with different markers (circles for the VLA, diamonds for ALMA, squares for AMI-LA, and pentagons for the ATA), and triangles mark $3 \sigma$ upper limits. We fitted these SEDs with the broken power-law function described in Eq. \ref{eq: bpl_function} (see detailed discussion in \S\ref{subsec: individual_analysis}).
\label{fig: radio_observations}}
\end{figure*}

Our $10$ GHz observation at $\Delta t = 88$ days did not reveal any detectable emission from the TDE with a luminosity upper limit of $\nu L_{\rm \nu} < 7.9 \times 10^{36} \, \rm erg \, s^{-1}$. At $\Delta t = 131$ days the $10$ GHz flux density is $\geq 40$ times larger than the initial upper limit, and the broadband SED indicates an optically thick emission, with a possible transition to the radio spectral peak around $17$ GHz at an observed peak luminosity of $\nu L_{\rm \nu} \approx 9 \times 10^{38} \, \rm erg \, s^{-1}$. The following broadband SEDs, up to $\Delta t = 173$ days, exhibit a fast decline of the peak flux density and of the peak frequency (see left panel of Fig.~\ref{fig: radio_observations}). In addition, the spectral shape of the SED evolved as well. The first ALMA observation, at $\Delta t = 139-141$ days, showed an optically thin spectral slope higher than $F_{\rm \nu} \sim \nu^{-1.2}$ (based on the detection at $97.5$ GHz and the upper limit at $203$ GHz). A later VLA+ALMA SED, obtained at $\Delta t = 151-154$ days, revealed an optically thin emission of $F_{\rm \nu} \sim \nu^{-0.55}$. These spectral slopes, together with the SED obtained at $\Delta t = 131$ days, imply the presence of an evolving spectral break frequency in the SED. This break frequency, typically associated with the cooling frequency $\nu_{\rm c}$ (in which the synchrotron emitting electrons quickly lose energy to radiation; see detailed discussion in \S\ref{sec:analysis}), evolved from $17 \, \rm{GHz} \, < \nu_{\rm c}< 97.5 \, \rm{GHz}$ to $\nu_{\rm c} \geq 97.5 \, \rm {GHz}$ between these two epochs.

At $\Delta t = 194$ days, a re-brightening of the radio emission was observed with AMI-LA at $15.5$ GHz. The evolution of this second radio flare is extremely peculiar. The observed peak luminosity is a factor of $\sim 3$ higher than the observed peak luminosity of the first radio flare. Furthermore, the temporal evolution of the broadband SEDs does not follow the temporal evolution of the first radio flare. The optically thin emission, seen in the mm-bands, showed a fast decline, and the spectral peak on the other hand rises while its peak frequency slowly moves to the lower-GHz bands.

In the first flare, assuming for now the optical discovery date as the reference time, the rise from the first upper limit ($\Delta t = 88$ days) to the first radio detection ($\Delta t = 131$ days) is at least as fast as $F_{\rm \nu} \sim t^{9}$. Then, the radio spectral peak declines as $F_{\rm p} \sim t^{-6}$, and the peak frequency also declines quickly, $\nu_{\rm p} \sim t^{-5}$. For the second flare, the detailed $15.5$ GHz light curve we obtained exhibits an extremely fast evolution of $F_{\rm 15.5 \, GHz} \sim t^{18}$, which then changes to $F_{\rm 15.5 \, GHz} \sim t^{4}$. Furthermore, the optically thin emission of the second flare, observed in the mm-bands, revealed an unprecedented fast decline of $F_{\rm \nu} \sim t^{-12}$ (see Fig. \ref{fig: light_curves}).

About $40\%$ of the optically-discovered TDEs show a late-time brightening of the radio emission \citep{Cendes_2024}. Naturally, as these TDEs rise at late times, a fast evolution of their radio emission with reference to their optical discovery date is expected. More specifically, when fitting the rise or the decay of the radio light-curve with a temporal power-law, $F_{\rm \nu} \sim t^{\alpha}$, a power-law index of $\left| \alpha \right| > 3$ is achieved. For example, the radio spectral peak of AT\,2018hyz evolved as $F_{\rm p} \simeq t^5$ \citep{Cendes_2022}. However, even compared to other radio-bright TDEs, the temporal evolution of the radio emission observed for AT\,2024tvd is the fastest ever seen in TDEs (see Fig. \ref{fig: radio_comparison}). 

The late-time emergence of the radio emission (for each flare separately) can mean that (a) a prompt sub-relativistic outflow was launched around the time of optical discovery, interacting with a complex medium, (b) a delayed sub-relativistic outflow was launched at $\Delta t > 0$ days, (c) a relativistic jet was launched around the time of optical discovery and the emission from this jet is entering our line of sight at late times, or (d) a relativistic jet was launched at late-times and we observed it either on- or off-axis. In \S\ref{sec:analysis}, we consider these different scenarios for the origin of the radio-emitting outflow(s). We discuss the astrophysical implications of our findings in \S\ref{sec: discussion}.

\section{Detailed modeling of the radio emission}
\label{sec:analysis}

Radio emission from transients is typically associated with synchrotron radiation generated in shocks as a result of fast-moving outflows. In the following subsections, we explore the possibilities of a prompt outflow (launched around the time of optical discovery), delayed outflows, and off-axis relativistic jets. In the standard picture of synchrotron radiation, at the shock front, electrons are accelerated to relativistic velocities and gain a power-law distribution of Lorentz factors, $N \left( \gamma \right) \, {\rm d} \gamma \propto \gamma^{-p} \, {\rm d} \gamma; \, \gamma \geq \gamma_{\rm m}$, where $\gamma_{\rm m}$ is the Lorentz factor of the lowest energy electrons, and magnetic fields are compressed and generated \citep{Sari_1998, chevalier_1998}. The presence of synchrotron self-absorption (SSA) results in a broken power-law synchrotron spectrum, where the exact spectral shape depends on the order of the minimal frequency $\nu_{\rm m}$ (which corresponds to $\gamma_{\rm m}$), the SSA frequency $\nu_{\rm sa}$, and the cooling frequency $\nu_{\rm c}$ which corresponds to $\gamma_{\rm c}$, above which electrons are losing significant energy to radiation (see \citealt{granot_2002} for the exact spectral shapes). 

So far, the radio-TDE literature has focused on synchrotron cooling as the dominant cooling mechanism. However, another important electron cooling mechanism is (external) inverse-Compton (IC) cooling in which thermal optical/UV photons are scattered to the X-ray band by the relativistic electrons at the shock front. This is most commonly used in the analysis of SNe \citep{chevalier_fransson_2006,margutti_2012,Nayana_2024}, and we consider it here for the first time in the context of TDEs (see detailed discussions in \S\ref{subsec: multi_epoch} and Appendix \ref{sec: prompt_sub_rel}). In the slow-cooling regime, where $\nu_{\rm m} < \nu_{\rm c}$, and assuming that $\nu_{\rm m} < \nu_{\rm sa}$, the radio spectral peak is at $\nu_{\rm sa}$, and a power-law index of $F_{\rm \nu} \sim \nu^{5/2}$ is expected for the optically thick regime above $\nu_{\rm m}$. In the optically thin regime, and for $\nu_{\rm sa} < \nu_{\rm c}$, the flux density is $F_{\rm \nu} \sim \nu^{-\left(p-1\right)/2}$ for $\nu < \nu_{\rm c}$, and $F_{\nu} \sim \nu^{-p/2}$ for $\nu > \nu_{\rm c}$. For $\nu_{\rm sa} > \nu_{\rm c}$ the optically thin regime is dominated by a single power-law, $F_{\rm \nu} \sim \nu^{-p/2}$. Finally, in cases where the material in front of the synchrotron generating shock is dense, external free-free absorption (FFA) can dominate in radio wavelengths and suppress the synchrotron emission. This external FFA was not considered so far for radio-emitting TDEs and we apply it for the first time in our analysis in \S\ref{subsec: multi_epoch}.

A common assumption is that a fraction, $\epsilon_{\rm e}$, of the post-shock energy density, $u_{\rm ps}= \frac{9}{8} \rho v_{\rm sh}^2$ (where $\rho$ is the density of the surrounding ambient medium, and $v_{\rm sh}$ is the velocity of the shock) is deposited in the relativistic electrons. A different fraction, $\epsilon_{\rm B}$, is deposited in the energy density of the magnetic fields, $u_{\rm B} = B^2 / 8\pi$. The assumption of equipartition implies that $\epsilon_{\rm e} = \epsilon_{\rm B}$. Here, the pre-factor of $9/8$ in the post-shock energy assumes it to be the thermal energy of a strong shock with an adiabatic index of $\gamma=5/3$, but we note that an order of unity correction to this energy is introduced for different scenarios (see the discussion in the appendix of \citealt{demarchi_2022}).

In \S\ref{subsec: individual_analysis}, we analyze individual broadband SEDs in the context of a synchrotron self-absorption model for different physical scenarios. While this form of analysis is very useful when inferring the physical parameters of the shock and its environment, it does not take into account the dynamics of the outflow. Therefore, with the purpose of also accounting for the shock-dynamics, in \S\ref{subsec: multi_epoch} we apply a time-dependent model to the radio observations of the first flare in the context of a non-relativistic outflow that interacts with a simple density profile.

\subsection{Analysis and modeling of individual SEDs}
\label{subsec: individual_analysis}

For a self-absorbed synchrotron spectrum, the radius of the radio-emitting region, and the magnetic-field strength can be inferred from the spectral peak of the radio emission. Here we use the Newtonian formalism introduced in \cite{chevalier_1998} since the inferred velocities in the spherical outflow scenario are sub-relativistic (see following sub-sections); however, we note that in case of a relativistic outflow the formalism introduced in \cite{BDNP13} is more suitable. If the spectral peak is dominated by $\nu_{\rm sa}$,  and $\nu_{\rm sa} < \nu_{\rm c}$, \cite{chevalier_1998} showed that the radius is given by
\begin{align}
    \label{eq: radius_chevalier}
    R = \left[ \frac{6 c_{\rm 6}^{p+5} F_{\rm \nu_{\rm sa}}^{p+6} d_{\rm l}^{2p+12}}{\alpha f (p-2) \pi^{p+5} c_{\rm 5}^{p+6} E_{\rm l}^{p-2}} \right]^{1/(2p+13)} \left( \frac{\nu_{\rm sa}}{2c_{\rm 1}} \right)^{-1}
\end{align}
and the magnetic field strength is
\begin{align}
    \label{eq: bfield_chevalier}
    B = \left[ \frac{36 \pi^{3} c_{\rm 5}}{\alpha^2 f^2 (p-2)^2 c_{\rm 6}^3 E_{\rm l}^{2(p-2)} F_{\rm \nu_{\rm sa}} d_{\rm l}^2} \right]^{2/(2p+13)} \left( \frac{\nu_{\rm sa}}{2c_{\rm 1}} \right)
\end{align}
where $F_{\rm \nu_{\rm sa}}$ is the peak flux density at the intersection between the optically thick and thin regimes of the synchrotron self-absorbed SED; $d_{\rm l}$ is the distance to the TDE; $f$ is the emission volume filling factor; $\alpha \equiv \epsilon_{\rm e}/\epsilon_{\rm B}$; $c_{\rm 1}$, $c_{\rm 5}$, and $c_{\rm 6}$ are constants provided in \cite{Pacholczyk_1970}; and $E_{\rm l}$ is the electron's rest-mass energy. Under the assumption of a strong shock, the number density of the surrounding, pre-shock, ambient medium is given by
\begin{align}
\label{eq: density_bfield}
    n_{\rm e} = \frac{1}{9 \pi \mu_{\rm e} m_{\rm p}} \epsilon_{\rm B}^{-1} B^2 v_{\rm sh}^{-2}
\end{align}
where $m_{\rm p}$ is the mass of a proton, and $\mu _{\rm e}$ is the mean molecular weight per electron (throughout our analysis we assume a fully ionized hydrogen composition, i.e., $\mu_{\rm e} = 1$).

To obtain the peak flux density and its frequency we adopt here a broken power-law function (see Eq. 1 in \citealt{granot_2002})
\begin{align}
    \label{eq: bpl_function}
    F_{\rm \nu} = F_{\rm p} \left[ \left( \frac{\nu}{\nu_{\rm p}} \right)^{-s\beta_{\rm 1}} + \left( \frac{\nu}{\nu_{\rm p}} \right)^{-s\beta_{\rm 2}} \right]^{-1/s}
\end{align}
where, $F_{\rm p}$ is the flux density at the intersection between the optically thick and thin regime, at $\nu_{\rm p}$. In the $\nu_{\rm sa} < \nu_{\rm c}$ scenario $\beta_{\rm 1} = 5/2$ and $\beta_{\rm 2} = -\left( p-1 \right)/2$. We fix\footnote{We note here that this value is given by \cite{granot_2002} for a wind-like density profile, i.e., $\rho \propto r^{-2}$. We later find that a steeper density profile is more suitable. However, the choice of the smoothing parameter does not change our conclusions.} the smoothing parameter, $s$, to $1.25 - 0.18p$.

We use \texttt{emcee} \citep{foreman_mackey_2013} to determine the best-fit parameters and infer their posterior distributions using flat priors. We use 200 walkers with 5,000 steps for each chain and discard the first 100 steps for burn-in. For both the first and the second radio flares, the broadband SEDs obtained with the VLA, ATA, and AMI-LA, do not probe the transition from the spectral peak to the optically thin regime to its full extent. Therefore, when fitting for the power-law index of the electrons, $p$, we combine VLA and ALMA observations separated by $2-3$ days. For the first flare, we infer $p$ based on VLA+ALMA observations at $\Delta t = 151-154$ days and apply it to observations at $\Delta t = 131$ days, $160$ days, and $173$ days. For the second flare, we infer $p$ based on VLA+ALMA observations at $\Delta t = 247-248$ days and apply it to observations at $\Delta t = 218$ days, $241-242$ days, and $284$ days. We note that while our fitting process constrains $F_{\rm p}$ and $\nu_{\rm p}$ at $\Delta t = 131$ days, this SED shows mostly optically thick emission, with only a slight transition to the radio spectral peak at the high frequencies in $13-17$ GHz. In addition, for the SED fitted at $\Delta t = 241-242$ days we used only ATA and AMI-LA data, and the ATA observations suffer from contamination because of its large beam. Therefore, we treat the inferred values from these epochs with caution. The results of these fits are reported in Table~\ref{tab: bpl_fits_parameters}, and we plot the best-fitting models in Fig.~\ref{fig: radio_observations}.

\begin{deluxetable}{cccc}[ht]
\tablecaption{Broken power-law fits to individual SEDs.}
\tablehead{
\colhead{$\Delta t$} & \colhead{$F_{\rm p}$}& \colhead{$\nu_{\rm p}$} & \colhead{$p$} \\
\colhead{$\rm \left[ Days \right]$} & \colhead{$\rm \left[ mJy \right]$} & \colhead{$\rm \left[ GHz \right]$}& \colhead{}}
\startdata 
$131$ & $2.08^{+0.28} _{-0.23}$ & $13.2^{+1.3} _{-1.1}$ & $2.14^{\dagger}$ \\
$151-154$ & $0.78^{+0.09} _{-0.07}$ & $5.71^{+0.52} _{-0.44}$ & $2.14^{+0.21} _{-0.14}$ \\
$160$ & $0.52^{+0.05} _{-0.04}$ & $4.28^{+0.38} _{-0.35}$ & $2.14^{\dagger}$ \\
$173$ & $0.41 \pm 0.05$ & $3.43^{+0.59} _{-0.53}$ & $2.14^{\dagger}$ \\
\hline
$218$ & $3.87 \pm 0.25$ & $6.03^{+0.47} _{-0.43}$ & $2.17^{\dagger}$ \\
$241-242$ & $5.86 \pm 0.49$ & $5.12^{+0.67} _{-0.58}$ & $2.17^{\dagger}$ \\
$247-248$ & $4.46 ^{+0.40} _{-0.36}$ & $4.40^{+0.37} _{-0.30}$ & $2.17 \pm 0.10$ \\
$284$ & $5.29 \pm 0.22$ & $2.80 \pm 0.20$ & $2.17^{\dagger}$ \\
\enddata
\tablecomments{Best-fitting parameters of the radio SEDs using Eq.~\ref{eq: bpl_function}. $\dagger$ marks observations for which the power-law index of the electron distribution is fixed (see discussion in \S\ref{subsec: individual_analysis}). The reported times and peak frequencies are in the frame of the observer.
\label{tab: bpl_fits_parameters}}
\end{deluxetable}

Based on the analysis above we infer $F_{\rm p}\sim t^{-6}$ and $\nu_{\rm p} \sim t^{-5}$ during the first flare and, $F_{\rm p}\sim t^{1.2}$ and $\nu_{\rm p} \sim t^{-3}$ during the late phases of evolution of the second flare. We next use these best-fitting parameters to derive the physical parameters of the shock. In the following subsections we explore different outflow scenarios using the results from the individual fits. In \S\ref{subsec: multi_epoch} we do a time-dependent analysis in the context of a sub-relativistic outflow for the first radio component only.

\subsubsection{A prompt, non-relativistic outflow}
\label{subsec: prompt_outflow}

\begin{figure*}[ht]
\centering
\includegraphics[width=0.495\linewidth]{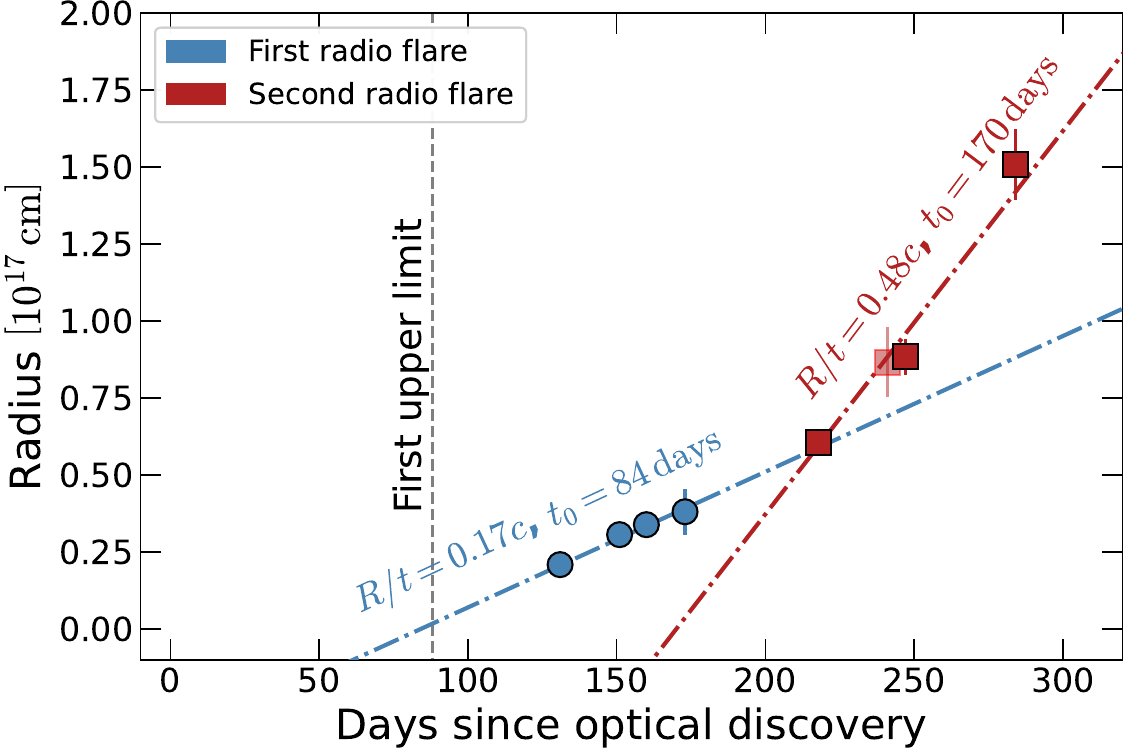}
\includegraphics[width=0.495\linewidth]{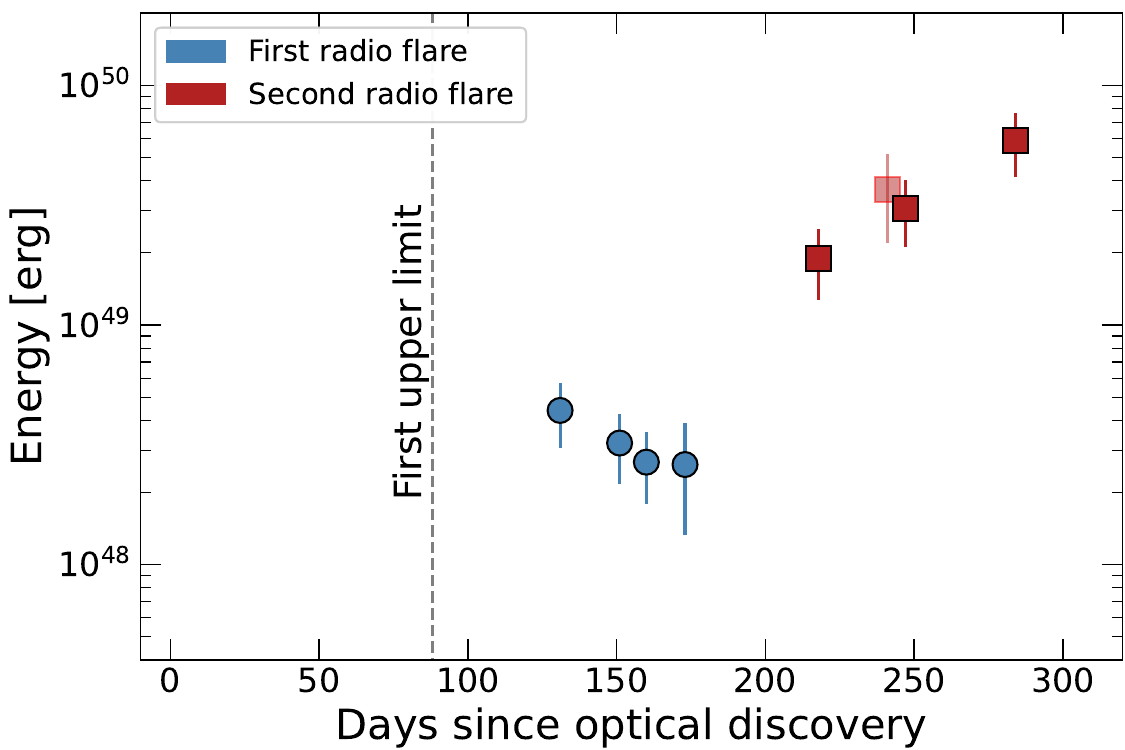}
\\
\includegraphics[width=0.7\linewidth]{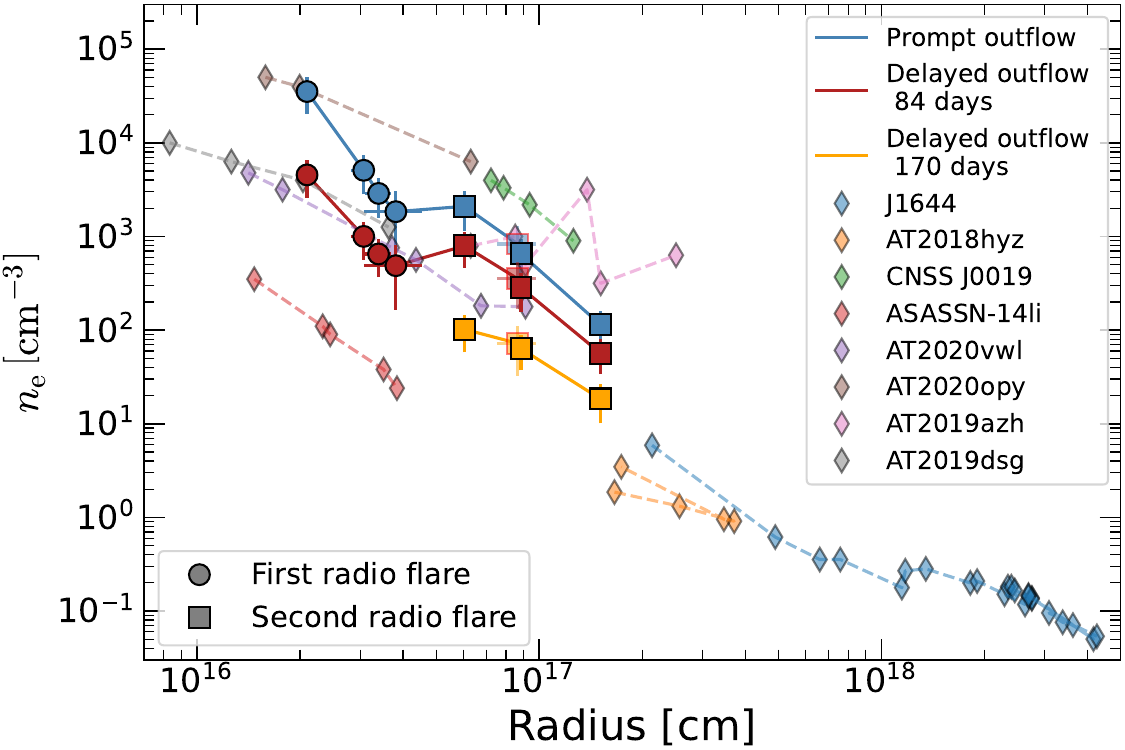}
\caption{Physical parameters inferred from the fits of the individual SEDs and non-relativistic equipartition analysis (see \S\ref{subsec: prompt_outflow} and \S\ref{subsec: delayed_outflow}). The top left panel shows the temporal evolution of the equipartition radius (filled circles for the first flare, filled squares for the second flare). Best-fitting $R\left( t \right)$ evolution for an assumed constant expansion of outflows that were launched at $t_{\rm launch} = 84^{+6} _{-25}$ days and $170 \pm 10$ days, are also plotted (dotted-dashed lines). The top right panel shows the equipartition energy. The bottom panel shows the density profile for an outflow that was launched at time of optical discovery (filled blue circles for the first flare; filled blue squares for the second flare) and for an outflow that was launched $84$ days after optical discovery (filled red circles for the first flare, filled red squares for the second flare). In addition we plot the density profile for a delayed outflow for the second flare only in yellow squares. Also plotted are the density profiles inferred for other radio-bright TDEs. The best-fitting parameters at $\Delta t \simeq 242$ days are marked with a lighter shade to emphasize our uncertainty in the flux measurements of the ATA due to its larger beam.
\label{fig: equipartition}}
\end{figure*}

First, we assume that (1) the outflow was launched around the time of optical discovery, (2) equipartition ($\epsilon_{\rm e} = \epsilon_{\rm B} = 0.1$), and (3) an emission filling factor, $f=0.5$, to derive (1) the equipartition radius (Eq.~\ref{eq: radius_chevalier}), (2) the magnetic field strength (Eq.~\ref{eq: bfield_chevalier}), (3) the density profile of the surrounding medium (Eq.~\ref{eq: density_bfield}), and (4) the energy of the event by assuming that the energy in the magnetic fields is a fraction of the post-shock energy 
\begin{align}
    \label{eq: energy_bfield_realtion}
    U_{\rm ps} = \epsilon_{\rm B}^{-1} \frac{B^2}{8 \pi} V ,
\end{align} 
where $V=\frac{4\pi}{3} f R^3$ is the volume of the emitting region. In Fig.~\ref{fig: equipartition} we present the results of this analysis and provide the inferred physical parameters in Table ~\ref{tab: equipartition_parameters}. 

\begin{deluxetable*}{ccccccc}[ht]
\tablecaption{Physical parameters from a non-relativistic equipartition analysis of individual SED fits.}
\tablehead{
\colhead{} & \colhead{$\Delta t$} & \colhead{$R$} & \colhead{$B$} &
\colhead{$\beta$} & \colhead{$n_e$} & \colhead{$U_{\rm ps}$} \\
\colhead{} & \colhead{$\rm \left[ Days \right]$} & \colhead{$\rm \left[ 10^{16} \, cm \right]$} & \colhead{$\rm \left[ G \right]$} & \colhead{} & \colhead{$\rm \left[ cm^{-3} \right]$} & \colhead{$\rm \left[ erg \right]$}}
\startdata 
& & & First radio flare & & \\
\hline
$t_{\rm launch} = 0$ days & $131$ & $2.10^{+0.15} _{-0.12}$ & $0.76^{+0.10} _{-0.07}$ & $0.062^{+0.005} _{-0.003}$ & $\left( 4.0^{+1.6} _{-0.9} \right) \times 10^{4}$ & $\left( 4.4^{+1.3} _{-0.7} \right) \times 10^{48}$ \\
& $151-154$ & $3.07 ^{+0.25} _{-0.22}$ & $0.37^{+0.05} _{-0.03}$ & $0.078^{+0.007} _{-0.006}$ & $\left( 5.8 ^{+2.5} _{-1.5} \right) \times 10^{3}$ & $\left( 3.2 ^{+0.9} _{-0.5} \right) \times 10^{48}$ \\
& $160$ & $3.39 ^{+0.31} _{-0.27}$ & $0.29^{+0.04} _{-0.03}$ & $0.082^{+0.008} _{-0.007}$ & $\left( 3.3 ^{+1.5} _{-0.9} \right) \times 10^{3}$ & $\left( 2.7 ^{+0.8} _{-0.5} \right) \times 10^{48}$ \\
& $173$ & $3.81 ^{+0.72} _{-0.59}$ & $0.24^{+0.05} _{-0.04}$ & $0.085^{+0.016} _{-0.013}$ & $\left( 2.1 ^{+2.0} _{-1.0} \right) \times 10^{3}$ & $\left( 2.6 ^{+1.2} _{-0.8} \right) \times 10^{48}$ \\
\hline
$t_{\rm launch} = 84$ days & $131$ & $-$ & $-$ & $0.17 \pm 0.01$ & $\left( 5.1^{+2.0} _{-1.2} \right) \times 10^{3}$ & $-$ \\
& $151$ & $-$ & $-$ & $0.18 \pm 0.01$ & $\left( 1.1 ^{+0.5} _{-0.3} \right) \times 10^{3}$ & $-$ \\
& $160$ & $-$ & $-$ & $0.17 \pm 0.02$ & $\left( 7.4 ^{+3.5} _{-2.1} \right) \times 10^{2}$ & $-$ \\
& $173$ & $-$ & $-$ & $0.17 \pm 0.03$ & $\left( 5.6 ^{+5.5} _{-2.7} \right) \times 10^{2}$ & $-$ \\
\hline
& & & Second radio flare & & \\
\hline
$t_{\rm launch} = 0$ days & $218$ & $6.05 ^{+0.42} _{-0.40}$ & $0.32^{+0.04} _{-0.03}$ & $0.107^{+0.008} _{-0.007}$ & $\left( 2.4 ^{+1.0} _{-0.6} \right) \times 10^{3}$ & $\left( 1.9 ^{+0.5} _{-0.3} \right) \times 10^{49}$ \\
& $241-242$ & $8.66^{+1.16} _{-1.07}$ & $0.26^{+0.05} _{-0.03}$ & $0.14 \pm 0.02$ & $\left( 9.6 ^{+7.0} _{-3.8} \right) \times 10^{2}$ & $\left( 3.7 ^{+1.4} _{-0.9} \right) \times 10^{49}$ \\
& $247-248$ & $8.85^{+0.59} _{-0.55}$ & $0.23^{+0.03} _{-0.02}$ & $0.14 \pm 0.01$ & $\left( 7.5 ^{+3.0} _{-1.9} \right) \times 10^{2}$ & $\left( 3.1 ^{+0.8} _{-0.5} \right) \times 10^{49}$ \\
& $284$ & $15.1 \pm 1.1$ & $0.14 \pm 0.02$ & $0.21 \pm 0.02$ & $\left( 1.3 ^{+0.5} _{-0.3}
\right) \times 10^{2}$ & $\left( 5.9^{+1.4} _{-0.9} \right) \times 10^{49}$ \\
\hline
$t_{\rm launch} = 84$ days & $218$ & $-$ & $-$ & $0.17 \pm 0.01$ & $\left( 9.1 ^{+3.7} _{-2.3} \right) \times 10^{2}$ & $-$ \\
& $241-242$ & $-$ & $-$ & $0.21 \pm 0.03$ & $\left( 4.1 ^{+3.0} _{-1.6} \right) \times 10^{2}$ & $-$ \\
& $247-248$ & $-$ & $-$ & $0.21 \pm 0.01$ & $\left( 3.3 ^{+1.3} _{-0.8} \right) \times 10^{2}$ & $-$ \\
& $284$ & $-$ & $-$ & $0.29 \pm 0.02$ & $65^{+26} _{-17}$ & $-$ \\
\hline
$t_{\rm launch} = 170$ days & $218$ & $-$ & $-$ & $0.49 \pm 0.03$ & $\left( 1.3 ^{+0.5} _{-0.3} \right) \times 10^{2}$ & $-$ \\
& $241-242$ & $-$ & $-$ & $0.47 \pm 0.06$ & $83^{+61} _{-33}$ & $-$ \\
& $247-248$ & $-$ & $-$ & $0.44 \pm 0.03$ & $73^{+29} _{-19}$ & $-$ \\
& $284$ & $-$ & $-$ & $0.51 \pm 0.04$ & $21^{+9} _{-6}$ & $-$ \\
\enddata
\tablecomments{The physical parameters of the outflows and their environment based on the broken power-law fits to the individual radio SEDs (\S\ref{subsec: individual_analysis}, \S\ref{subsec: prompt_outflow}, and \S\ref{subsec: delayed_outflow}). The radius and magnetic field strength are calculated using Eq. \ref{eq: radius_chevalier} and \ref{eq: bfield_chevalier}, respectively. The normalized velocity, $\beta$, is defined by $R/\left( \Delta t - t_{\rm launch }\right) \equiv \beta c$. The number density is calculated using Eq. \ref{eq: density_bfield}. The last column is of the post-shock equipartition energy and it is calculated using Eq.~\ref{eq: energy_bfield_realtion}. Since the radius, magnetic field, and equipartition energy are not time-dependent, we do not repeat these values for different $t_{\rm launch}$ (this is marked with "$-$"). These reported values assume $\epsilon_{\rm e} = \epsilon_{\rm B} = 0.1$, and $f=0.5$.
\label{tab: equipartition_parameters}}
\end{deluxetable*}

We find that the inferred velocities are sub-relativistic for both flares (see Fig.~\ref{fig: equipartition}). Interestingly, the radius derived from the first SED of the second radio flare matches the radius evolution during the first radio flare. This suggests that both radio flares are a result of the same outflow. However, we cannot estimate the radius of the emitting region during the second radio flare before the first broadband SED. Therefore, the fact that the initial radius estimation during the second radio flare matches the observed evolution during the first radio flare might also be a coincidence. These results also imply significant acceleration, $R \sim t^{2.5}$ and $R \sim t^{3.5}$ for the first and second radio components, respectively. The initial equipartition energy of the first flare is $\sim 4 \times 10^{48} \, \rm erg$ at $\Delta t = 131$ days, and then declines with time. The energy of the second flare increases with time, reaching $\sim 6 \times 10^{49} \, \rm erg$ at $\Delta t = 284$ days, a factor of $\simeq 10$ higher than the first flare (see  Fig. \ref{fig: equipartition}). This overall behavior motivates the scenarios of two distinct delayed outflows and/or off-axis relativistic jets.

The density of the surrounding medium depends on the velocity of the shock (Eq.~\ref{eq: density_bfield}). Here we estimate the velocity for each epoch by $v_{\rm sh} = R/\Delta t$ since we are conducting a time-independent analysis (see \S\ref{subsec: multi_epoch} for a time-dependent analysis). For the first flare, we find $n_{\rm e} = \left( 4.0^{+1.6} _{-0.9} \right) \times 10^{4} \, \rm cm^{-3}$ at $R = \left( 2.10^{+0.15} _{-0.12} \right) \times 10^{16} \, \rm cm$, and a very steep density profile, $n_{\rm e} \sim r^{-5}$ (see bottom panel of Fig.~\ref{fig: equipartition}). However, extrapolating this density profile to the time of the first radio non-detection ($\Delta t = 88$ days), and assuming a similar velocity to the velocity measured at $\Delta t = 131$ days, results in a $10$ GHz flux density higher than the radio upper limit that was obtained at this time, even when accounting for external FFA by the dense medium in-front of the radio emitting shock.

Therefore, this density profile cannot explain the entire radio emission up to $\Delta t = 173$ days. One possibility is that there is a change in the density profile at $R < 2 \times 10^{16}$ cm. In the case of a density cavity, the emission at $\Delta t = 88$ days is too faint to be detected. On the other hand, if there is an over-density at these small radii, the radio emission can be suppressed by free--free absorption by the dense medium in front of the shock. We discuss the densities needed to suppress the emission by FFA in \S\ref{subsec: multi_epoch} and Appendix \S\ref{sec: prompt_sub_rel}. Finally, the same analysis for the second radio flare (assuming a prompt outflow) requires a change in the density profile around $\left( 4-6 \right) \times 10^{16} \, \rm cm$, and additional acceleration.

\subsubsection{Delayed, non-relativistic outflows}
\label{subsec: delayed_outflow}

From the analysis above, $R \left( t \right)$ is neither constant nor decelerating for $t_{\rm launch} = 0$ days, where we define $t_{\rm launch}$ as the launch time (in days from optical discovery) of the radio-emitting outflow. While one possibility for this observed behavior is shock acceleration, e.g., due to the propagation of the shock in a steep, $\rho \sim r^{-s}$, with $s > 3$ density profile \citep{Waxman_1993}, it is also possible that one, or two, outflow(s) were launched at $t_{\rm launch} > 0$ days, resulting in an apparent shock acceleration. To estimate the outflow launch time we fit the equipartition radius with a function of the form $R \left( \Delta t \right) = \tilde{R} \left( \frac{\Delta t - t_{\rm launch}}{\Delta t_{\rm 0} - t_{\rm launch}}\right)$, which assumes a constant outflow expansion velocity of $v_{\rm sh} \equiv \frac{\tilde{R}}{\Delta t_0 - t_{\rm launch}}$. We set $\Delta t_{\rm 0} \equiv 131$ and $247$ days for the first and second flare, respectively. $\tilde{R}$ and $t_{\rm launch}$ are free parameters. \texttt{emcee} is used to estimate the posteriors of the best-fitting parameters. For the first flare the outflow launching time is $t_{\rm launch} = 84^{+6} _{-25}$ days, and the radius at $\Delta t_{\rm 0} = 131$ days is $\tilde{R} = \left( 2.1 \pm 0.1\right) \times 10^{16} \, \rm cm$; for the second flare $t_{\rm launch} = 170 \pm 10$ days and $\tilde{R} = \left( 1.0 \pm 0.1 \right) \times 10^{17} \, \rm cm$.

The equipartition radius and magnetic field strength do not depend on $t_{\rm launch}$. However, since $n_{\rm e} \propto B^{2} v_{\rm sh}^{-2}$ (Eq.~\ref{eq: density_bfield}), and we infer higher velocities, the assumption of a delayed outflow results in lower number density than in the case of $t_{\rm launch} = 0$ days. While the required densities are lower, the need for a steep $n_{\rm e} \left( r\right)$ profile stays ($n_{\rm e} \sim r^{-4}$ here vs. $n_{\rm e} \sim r^{-5}$ in the prompt outflow scenario). In addition, if the first and second flares are manifestations of the same outflow (i.e., they share the same $t_{\rm launch}$) a density enhancement is still needed, and a shallower, $n_{\rm e} \sim r^{-3}$, density profile is inferred at radii $\gtrsim 6 \times 10^{16} \, \rm cm$. If the second radio flare is a result of an additional delayed outflow with $t_{\rm launch} = 170$ days, then we find a mildly-relativistic shock with $v_{\rm sh} \simeq 0.5 c$, and no need for density enhancement at large radii. Since the velocity we infer is still only mildly-relativistic, the Newtonian formalism is sufficient, and we do not use the relativistic formalism discussed in \cite{BDNP13}. Finally, we note that the lack of emission at early times, in both radio flares, is purely due to the delayed launching of the outflow(s), and we do not need to invoke FFA. We provide the results of this analysis in Table \ref{tab: equipartition_parameters}, and plot the different density profiles in the bottom panel of Fig.~\ref{fig: equipartition}.

\subsubsection{Relativistic jets observed off-axis}
\label{subsec: off_axis_jet}

\begin{figure}[ht]
\includegraphics[width=\linewidth]{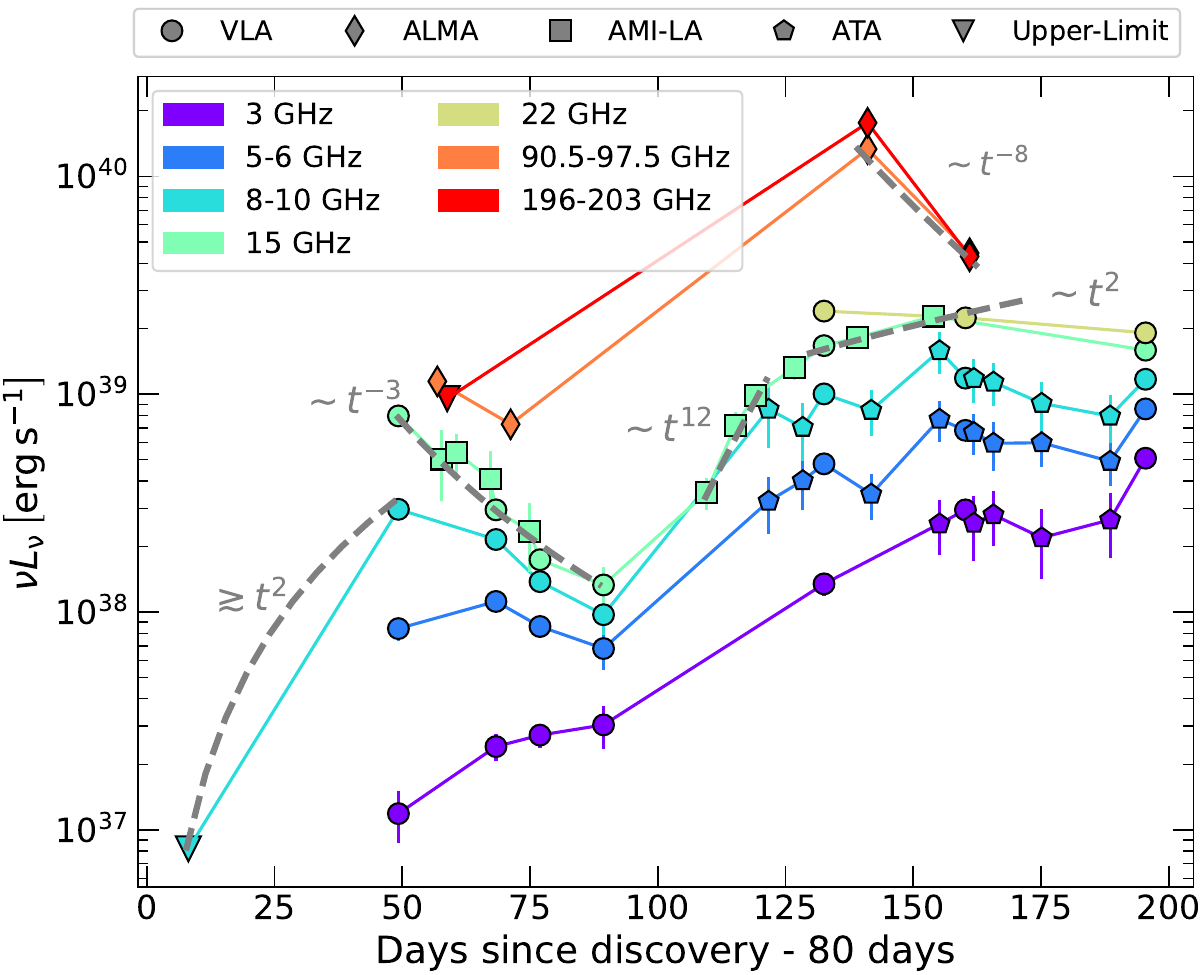}
\caption{Radio light curves of AT\,2024tvd as in Fig.~\ref{fig: light_curves} with power-law evolution for $t_{\rm launch} = 80$ days.
\label{fig: light_curves_dalayed}}
\end{figure}

Another scenario that can explain the emergence of radio emission at late-times in TDEs is a relativistic jet that is initially observed off-axis \citep{Giannios_2011, Matsumoto_2023, Beniamini_2023_J1644}. \cite{Matsumoto_2023} introduced a generalized equipartition formalism, demonstrating that a relativistic emitter observed off-axis can be disguised as an apparent on-axis Newtonian outflow. Forward modeling of the synchrotron emission suggested that the peculiar evolution of the radio emission from both Swift J1644+57 \citep{Beniamini_2023_J1644} and AT\,2018hyz \citep{Sfaradi_2024} can be a result of a relativistic off-axis jet.\footnote{An on-axis jet but with a wide angular structure carrying more energy than the jet core, can also produce a late-time rebrightening similar to that seen in Swift J1644+57 (\citealt{Mimica+15}).} In this context, \cite{Beniamini_2023_J1644} derived that the fastest possible observed rise is $F_{\rm \nu} \sim t^{10}$ (achieved for a flat density profile and a top-hat jet), and the fastest decline is $F_{\rm \nu} \sim t^{-p}$. These closure relations can explain the fast rise, $F_{\rm \nu} \gtrsim t^{9}$, observed in the first flare for $t_{\rm launch} = 0$ days. However, the extremely steep rise of the second flare ($F_{\rm \nu} \gtrsim t^{18}$), and the declines of both flares ($F_{\rm \nu} \sim t^{-6}$ and $t^{-12}$ for the first and second flare, respectively) are too fast for an off-axis jet launched promptly after discovery, thus requiring $t_{\rm launch} > 0$ days in both cases. We set $t_{\rm launch}$ based on the requirement that the emission declines as $F_{\rm \nu} \sim t^{-3}$, and note that this is a conservative approach as the emission is expected to follow $F_{\rm \nu} \sim t^{-p}$, and we observe $p \simeq 2.15$ for both flares. This approach results in $t_{\rm launch} \gtrsim 80$ days for the first flare, and $t_{\rm launch} \gtrsim 190$ days for the second flare. For reference, in Fig.~\ref{fig: light_curves_dalayed} we present the radio light curves with the temporal indices for the case of $t_{\rm launch} \simeq 80$ days.

We explore the possibility of an off-axis jet model by generalizing the method devised by \cite{Matsumoto_2023}. In this model, the observables are transformed from the rest-frame quantities by the relativistic Doppler factor $\delta_{\rm D} \equiv \frac{1}{\Gamma \left( 1- \beta \, \cos \theta\right)}$, where $\Gamma \equiv \frac{1}{\sqrt{1-\beta^2}}$, and $\theta$ is the angle between the direction of motion of the jet and the line of sight of the observer. \cite{BDNP13} generalized equipartition analysis by accounting for electrons that radiate at $\nu_{\rm m}$. This introduces additional factors that arise from the fact that the energy of the electrons is higher (see discussion in their \S4.2.1). We generalize the \cite{Matsumoto_2023} derivation by accounting for these electrons, and for a general power-law index of the relativistic electrons, $p$. We find (a full derivation is in Appendix \ref{sec: off_axis_derivation}):
\begin{align}
    \label{eq: off_axis_radius} R_{\rm eq} = & \left( 10^{17} \, {\rm cm}\right) \left[ 159 \times\left[520\right]^{p-1} \right]^{\frac{1}{2p+13}} \times \\ 
    & \nonumber F_{\rm p, mJy}^{\frac{p+6}{2p+13}} d_{\rm L, 28}^{\frac{2\left( p+6 \right)}{2p+13}} \nu_{\rm p, 10}^{-1} \left( 1 + z \right)^{-\frac{3p + 19}{2p+13}} \epsilon^{\frac{1}{2p+13}} \times \\
    & \nonumber \gamma_{\rm m}^{\frac{2-p}{2p+13}} f_{\rm A}^{-\frac{p + 5}{2p+13}} f_{\rm V}^{-\frac{1}{2p+13}} \Gamma \delta_{\rm D}^{- \frac{p + 5}{2p+13}} \xi^{\frac{1}{2p+13}}, 
    \\
    \nonumber \\
    \label{eq: off_axis_energy} E_{\rm eq} = & \left( 3.36 \times 10^{48} \, {\rm erg}\right) \left[ 159 \right]^{-\frac{2 \left( p+1\right)}{2p+13}} \left[520\right]^{\frac{11\left(p-1\right)}{2p+13}} \times \\
    & \nonumber F_{\rm p, mJy}^{\frac{3p+14}{2p+13}} d_{\rm L, 28}^{\frac{2\left( 3p+14 \right)}{2p+13}} \nu_{\rm p, 10}^{-1} \left( 1 + z \right)^{-\frac{5p + 27}{2p+13}} \times
    \\
    \nonumber & \left[ 1 + \left( \frac{2p+2}{11} \right) \epsilon \right] \epsilon^{\frac{-2\left(p+1\right)}{2p+13}} \gamma_{\rm m}^{\frac{11 \left(2-p\right)}{2p+13}} f_{\rm A}^{-\frac{3\left(p + 1\right)}{2p+13}} \times \\
    & \nonumber f_{\rm V}^{\frac{2\left(p+1\right)}{2p+13}} \Gamma \delta_{\rm D}^{- \frac{7p + 29}{2p+13}} \xi^{\frac{11}{2p+13}}, 
\end{align}
where $F_{\rm p,mJy}$ is the peak flux density in mJy; $d_{\rm L,28}$ is the luminosity distance normalized to $10^{28} \, \rm cm$; $\nu_{\rm p,10}$ is the observed peak frequency normalized to $10 \, \rm GHz$; $z$ is the redshift to the source; $f_{\rm A}$ and $f_{\rm V}$ are the area and volume filling factors, respectively; $\epsilon \equiv \left( \epsilon_{\rm B}/\epsilon_{\rm e} \right) /\left( \frac{2p+2}{11} \right)$, and $\xi \equiv 1+\epsilon_{\rm e}^{-1}$. For this relativistic case we consider $\gamma_{\rm m} = \max \left[2, \epsilon_{\rm e} 
\frac{p-2}{p-1} \frac{m_{\rm p}}{m_{\rm e}} \left( \Gamma - 1 \right) \right]$ as the minimal Lorentz factor of the radiating electrons.

\begin{figure*}[ht]
\includegraphics[width=\linewidth]{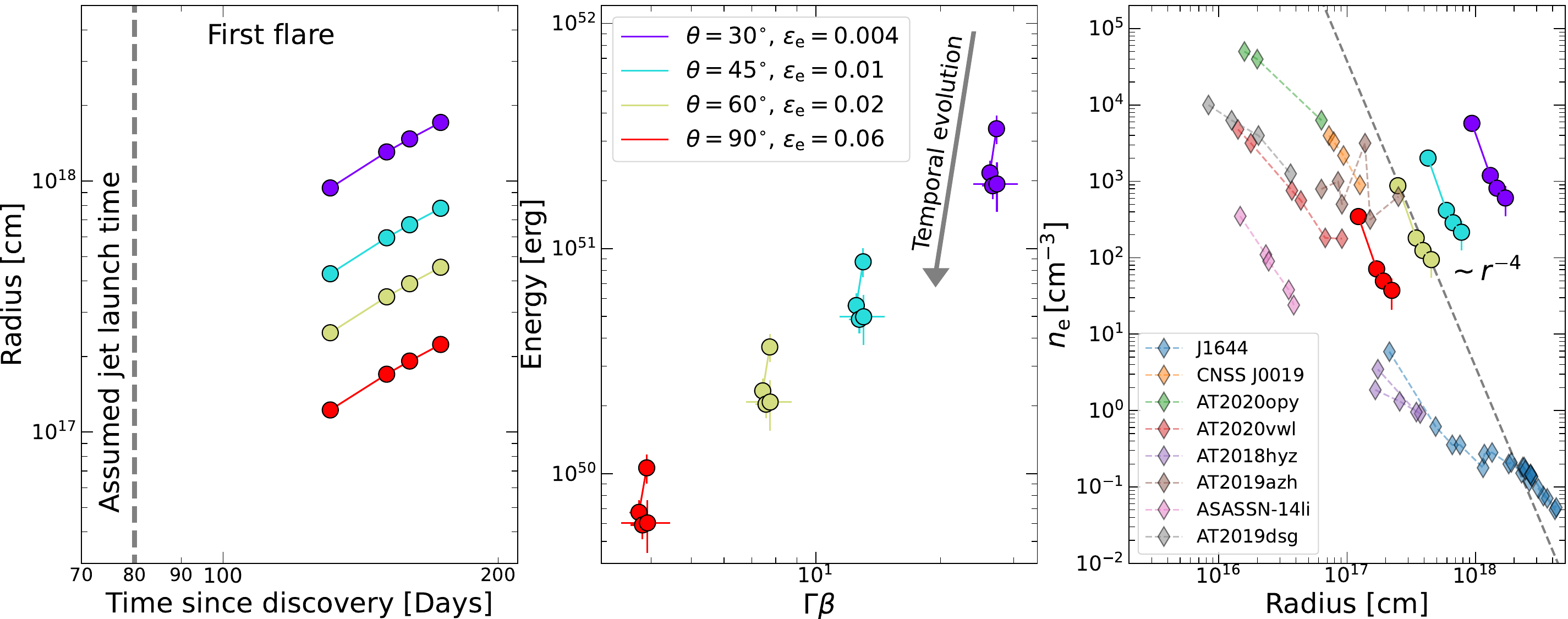}
\\
\includegraphics[width=\linewidth]{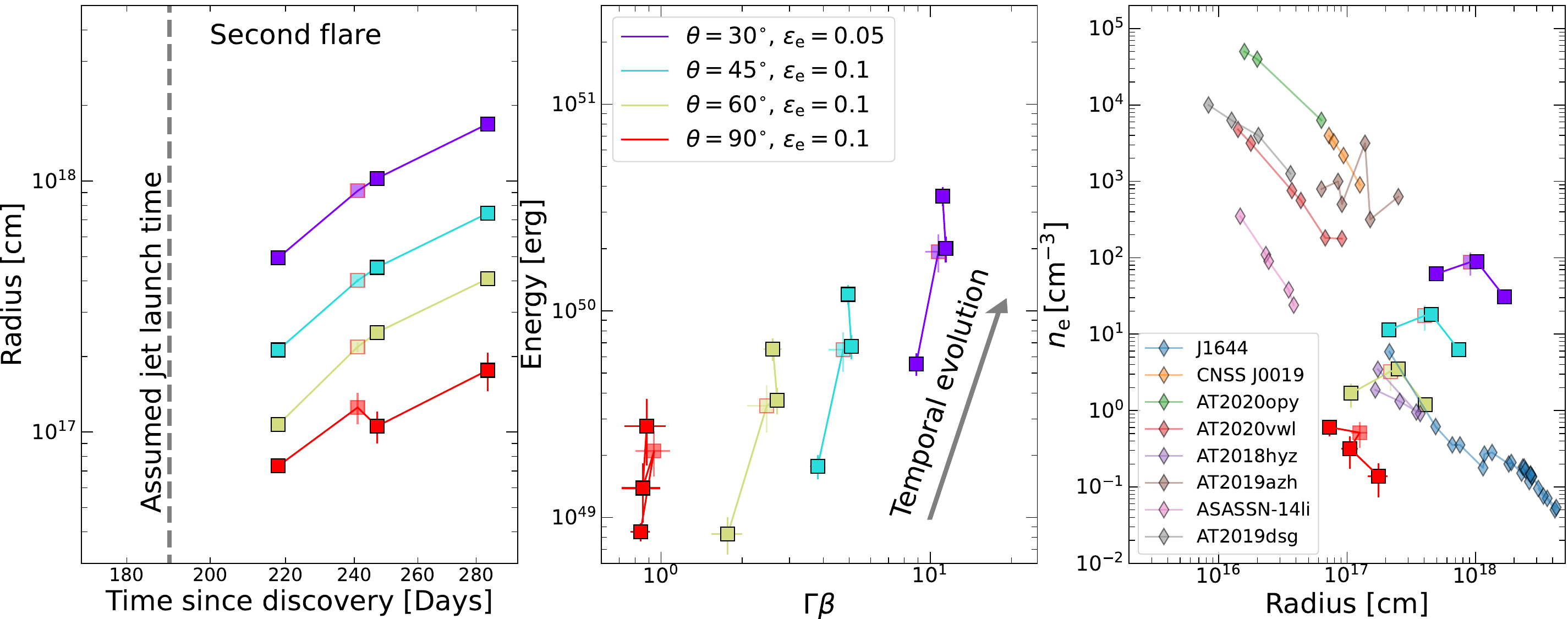}
\caption{Physical parameters from the generalized equipartition analysis of a relativistic off-axis jet. The top panels show the results for the first radio flare, assuming $t_{\rm launch} = 80$ days, and the bottom panels show the results for the second radio flare for $t_{\rm launch} = 190$ days. The left panels show the evolution of the radius with time, the middle panels show the equipartition energy vs. proper velocity (we mark the temporal evolution with an arrow), and the right panels display the inferred density profile compared to other radio-bright TDEs. We show the results of this analysis for an observer angle of $\theta = 30^{\circ}$, $45^{\circ}$, $60^{\circ}$, and $90^{\circ}$. For each observing angle we use the largest $\epsilon_{\rm e}$ (up to $\epsilon_{\rm e} = 0.1$) that satisfies the relation $\nu_{\rm m} \leq 1.5$ GHz in all of our epochs (see discussion in \S\ref{subsec: off_axis_jet}).
\label{fig: off_axis_equipartition}}
\end{figure*}

\begin{deluxetable*}{ccccccc}[ht]
\tablecaption{Physical parameters derived from the relativistic off-axis equipartition analysis of individual SED fits (\S\ref{subsec: individual_analysis}).}
\tablehead{
\colhead{} & \colhead{$\Delta t$} & \colhead{$R_{\rm eq}$} &
\colhead{$E_{\rm eq}$} &
\colhead{$\Gamma \beta$} &
\colhead{$n_e$} \\
\colhead{} & \colhead{$\rm \left[ Days \right]$} & \colhead{$\rm \left[ 10^{17} \, cm \right]$} & \colhead{$\rm \left[ erg \right]$} & \colhead{} & \colhead{$\rm \left[ cm^{-3} \right]$}}
\startdata
& & & First radio flare & - $t_{\rm launch} \equiv 80$ days & & \\
\hline
$\theta = 45^{\circ}$ & $131$ & $4.27 \pm 0.01$ & $\left( 8.7 \pm 1.0 \right) \times 10^{50}$ & $13.0 \pm 0.4$ & $\left( 2.0 \pm 0.4\right) \times 10^{3}$ & \\
$\epsilon_{\rm e} = \epsilon_{\rm B} = 0.01$ & $151-154$ & $5.94 \pm 0.01$ & $\left( 5.6 \pm 0.7 \right) \times 10^{50}$ & $12.5 \pm 0.6$ & $\left( 4.2^{+0.9} _{-0.7}\right) \times 10^2$  \\
& $160$ & $6.70 \pm 0.01$ & $\left( 4.9^{+0.7} _{-0.6}\right) \times 10^{50}$ & $12.7 \pm 0.7$ & $\left( 2.8^{+0.6} _{-0.5}\right) \times 10^{2}$ \\
& $173$ & $7.79 \pm 0.02$ & $\left( 5.0 \pm 1.0 \right) \times 10^{50}$ & $13.0 \pm 1.6$ & $\left( 2.2^{+1.0} _{-0.7} \right) \times 10^{2}$ \\
\hline
$\theta = 90^{\circ}$ & $131$ & $1.23 \pm 0.01$ & $\left( 1.1 \pm 0.2 \right) \times 10^{50}$ & $3.90 \pm 0.15$ & $\left( 3.5^{+0.7} _{-0.6}\right) \times 10^{2}$ \\
$\epsilon_{\rm e} = \epsilon_{\rm B} = 0.06$ & $151-154$ & $1.70 \pm 0.01$ & $\left( 6.7 \pm 0.9 \right) \times 10^{49}$ & $3.7 \pm 0.2$ & $71^{+15} _{-13}$ \\
& $160$ & $1.92 \pm 0.01$ & $\left( 5.9 \pm 0.8 \right) \times 10^{49}$ & $3.8 \pm 0.2$ & $49^{+11} _{-10}$ \\
& $173$ & $2.23 \pm 0.02$ & $\left( 6.0 \pm 1.5 \right) \times 10^{49}$ & $3.9 \pm 0.5$ & $37^{+17} _{-13}$ \\
\hline
& & & Second radio flare & - $t_{\rm launch} \equiv 190$ days & \\
\hline
$\theta = 45^{\circ}$ & $218$ & $2.12 \pm 0.03$ & $\left(1.8 \pm 0.2 \right) \times 10^{49}$ & $3.8 \pm 0.2$ & $11 \pm 2$ \\
$\epsilon_{\rm e} = \epsilon_{\rm B} = 0.1$ & $241-242$ & $4.0 \pm 0.07$ & $\left(6.5 \pm 1.0 \right) \times 10^{49}$ & $4.7 \pm 0.6$ & $17\pm 6 $ \\
& $247-248$ & $4.53 \pm 0.03$ & $\left( 6.7^{+1.0} _{-0.8} \right) \times 10^{49}$ & $5.1 \pm 0.3$ & $18^{+4} _{-3}$ \\
& $284$ & $7.44 \pm 0.06$ & $\left( 1.2 \pm 0.1\right) \times 10^{50}$ & $4.9 \pm 0.3$ & $6 \pm 1$ \\
\hline
$\theta = 90^{\circ}$ & $218$ & $0.73 \pm 0.01$ & $\left( 8.5 \pm 0.8 \right) \times 10^{48}$ & $0.84 \pm 0.07$ & $0.6 \pm 0.1$ \\
$\epsilon_{\rm e} = \epsilon_{\rm B} = 0.1$ & $241-242$ & $1.35^{+0.04} _{-0.34}$ & $\left( 2.1^{+0.4} _{-0.8} \right) \times 10^{49}$ & $0.94 \pm 0.15$ & $0.5 \pm 0.2$ \\
& $247-248$ & $1.05^{+0.34} _{-0.05}$ & $\left( 1.4^{+0.8} _{-0.2} \right) \times 10^{49}$ & $0.8^{+0.2} _{-0.1}$ & $0.3^{+0.2} _{-0.1}$ \\
& $284$ & $1.7^{+0.6} _{-0.1}$ & $\left( 2.8^{+1.6} _{-0.6} \right) \times 10^{49}$ & $0.9 \pm 0.2$ & $0.14^{+0.07} _{-0.05}$ \\
\enddata
\tablecomments{Physical parameters of a relativistic off-axis jet, and its environment, based on the broken power-law fits of individual radio SEDs, and the generalized equipartition analysis described in \S\ref{subsec: off_axis_jet} and Appendix \ref{sec: off_axis_derivation}. We report here the values for observing angles of $45^{\circ}$ and $90^{\circ}$ (Fig.~\ref{fig: off_axis_equipartition} also shows the results for $\theta = 30^{\circ}$ and $60^{\circ}$). We set different launch time of the jet for the first and second flare based on the requirement that the decline of the radio emission will not exceed $F_{\rm \nu} \sim t^{-3}$. We report here the largest value of (up to $\epsilon_{\rm e} = 0.1$) that satisfies the condition $\nu_{\rm m} < 1.5$ GHz (see details in \S\ref{subsec: off_axis_jet}).
\label{tab: equipartition_parameters_off}}
\end{deluxetable*}

To derive the physical parameters of an off-axis jet we follow the same procedure outlined in \cite{Matsumoto_2024} and \cite{Christy_2024}. The radius, $R$, and the time of observation since the jet was launched, $t$, are related by the time measured with a photon emitted by a relativistically moving source with normalized velocity $\beta$ at an angle $\theta$
\begin{align}
\label{eq: jet_temporal_evolution}
    t = \frac{\left( 1 + z \right) R}{c \beta} \left( 1 - \beta \cos \theta \right).
\end{align}
We then solve numerically for the radius in Eq. \ref{eq: off_axis_radius} and \ref{eq: jet_temporal_evolution} for the following viewing angles, $\theta = 30^{\circ}, \, 45^{\circ}, \, 60^{\circ}$, and $90^{\circ}$. For the surrounding density profile we assume a strong shock, i.e., $n_{\rm e} = \frac{N_{\rm e}}{4V}$, where $N_{\rm e}$ is the number of emitting electrons (Eq.~\ref{eq: Ne_off}), and $V \equiv f_{\rm V} \pi R_{\rm eq}^2/\Gamma^4$ is the emitting volume. We assume $f_{\rm A} = f_{\rm V} = 1$, which accounts only for energy within $1/\Gamma$ from our line of sight. 

We note that for an assumed $\epsilon_{\rm e} = 0.1$ the large Lorentz factors we infer for some of the viewing angles result in $\nu_{\rm sa} < \nu_{\rm m} , \nu_{\rm c}$ which is not consistent with our observations (we find a steeper spectral slope in the optically thick regime than all the cases discussed in \citealt{granot_2002} with $\nu_{\rm sa} < \nu_{\rm m}$). Our observations imply $\nu_{\rm m} \lesssim 1.5 \, {\rm GHz} < \nu_{\rm sa}$ and, in order to be self-consistent, we re-derive the generalized equipartition formalism introduced in \cite{Matsumoto_2023} for the correct frequency ordering (Appendix \ref{sec: off_axis_derivation}), and add the requirement of $\nu_{\rm m} \lesssim 1.5$ GHz. We note that this requirement is necessary as we do not observe the actual position of $\nu_{\rm m}$, and therefore assume that it is below our observing range. The inferred $\epsilon_{\rm e}$ are thus upper limits in this case. Finally, we also assume equipartition, $\epsilon_{\rm e} = \epsilon_{\rm B}$. The results of this analysis are shown in Fig.~\ref{fig: off_axis_equipartition} and summarized in Table~\ref{tab: equipartition_parameters_off}.

For the first radio flare we find that a relativistic jet launched $\sim 80$ days after optical discovery at $\theta \simeq 60^{\circ}-90^{\circ}$ can explain the observed emission with moderate Lorentz factors and density profiles similar to those seen in other radio bright TDEs. Smaller viewing angles require larger $\Gamma$ and density profiles: for $\theta = 30^{\circ}$ we find $\Gamma \sim 30$, and a density profile $\sim 3-4$ orders of magnitude larger than seen in Swift J1644+57 at radii $\gtrsim 10^{18} \, \rm cm$ (see top panels of Fig.~\ref{fig: off_axis_equipartition}). 

For the second radio flare we infer $t_{\rm launch} = 190$ days (based on the requirement of $F_{\rm \nu} \sim t^{-3}$ in the optically thin regime), which is only $\sim 4$ days before the first detection of the second radio flare. Therefore, any relativistic jet has to be either only slightly off-axis or only mildly-relativistic for us to see it that soon (see bottom panels of Fig.~\ref{fig: off_axis_equipartition}). Overall we find moderate Lorentz factors, $\Gamma \leq 10$, and densities, $n_{\rm e} \lesssim 100 \, \rm cm^{-3}$ up to $\sim 10^{18} \, \rm cm$, for all the viewing angles we have tested. By relaxing the requirement of a delayed jet for the second flare, and assuming a prompt off-axis jet, we infer $\Gamma > 10$ and $E_{\rm eq} \gtrsim 10^{51} \, \rm erg$ for $\theta < 60^{\circ}$. In this scenario, an extreme density profile of $\rho \sim r^{-15}$ is inferred. Therefore, we conclude that an off-axis relativistic jet launched around the time of optical discovery is disfavored.

\subsection{Multi-epoch time-dependent modeling of the first radio flare}
\label{subsec: multi_epoch}

We next refine our non-relativistic analysis by adopting a time-dependent model to fit the emission from the first radio flare only, and show that it can be explained by a shockwave interacting with a single power-law density profile. We attempted the same approach for the second flare. However, more epochs of observations are needed to constrain the evolution of the optically thin regime. We leave the modeling of the second radio flare beyond equipartition for future work.

We assume that the emission is from a shock wave traveling with roughly constant velocity, $r \left( \Delta t \right) = \beta_{\rm 0}c \left( \Delta t - t_{\rm launch}\right)$, and interacting with a single power-law density profile $n_{\rm e} \left( r \right) = n_{\rm 0} \left( r/r_{\rm 0} \right)^{-k}$, where $r_{\rm 0} \equiv \beta_{\rm 0} c \left(\Delta t_{\rm 0} - t_{\rm launch} \right)$, and we choose $\Delta t_{\rm 0} = 131$ days as our reference time (the time of the first radio detection)\footnote{Note that we find in our analysis $k>3$. This steep profile should lead to an acceleration of the shock wave \citep{Waxman_1993}. However, for simplicity, we do not account for this acceleration and assume constant expansion and leave this to future work.}. We use the model presented in Appendix \ref{sec: ssa_frequency} to self-consistently calculate the different synchrotron break frequencies, and specifically Eq. \ref{eq: fp_nup_multi} to calculate $F_{\rm p}$ and $\nu_{\rm p}$. We then use Eq. \ref{eq: bpl_function} to calculate the SED shape before and after the spectral peak. The shape parameter around the peak, $s$, is defined for each transition in \cite{granot_2002}. We use sharp breaks for other spectral breaks that are not around the peak, and follow the power-law indices discussed in \cite{granot_2002}. This results in a self-absorbed synchrotron spectrum $F_{\rm \nu , \, ssa} \left( t \right)$. We also calculate the free-free absorption optical depth as in \cite{Rybicki_lightman} (see also \citealt{Nayana_2024}):
\begin{align}
\label{eq: tau_ffa}
    \tau_{\rm FFA} \left( r, \nu \right) = \int^{\infty} _{r} 0.018 \times T_{\rm e} ^{-3/2} Z^{2} \nu^{-2} g_{\rm ff} n_{\rm e}^2 \left( r' \right) dr'
\end{align}
where $T_{\rm e}$ is the temperature of the electrons in the surrounding medium and we use a Gaunt factor $g_{\rm ff} = 5$, a charge of $Z=1$, and assume full ionization of the medium. We then calculate $\nu_{\rm ff}$ by setting $\tau_{\rm FFA} \left( r, \nu_{\rm ff} \right) \equiv 1$. The resulting spectrum is given by $F_{\rm \nu} \left( t \right) = F_{\rm \nu \, ssa} \left( t \right) \times e^{-\tau_{\rm FFA}} = F_{\rm \nu \, ssa} \left( t \right) \times e^{-\left( \nu/\nu_{\rm ff} \right)^{-2}}$. Finally, we also account for the effects of IC cooling by thermal optical/UV photons. IC cooling effects are typically neglected in the TDE literature. However, they are relevant for the first radio flare, as the thermal emission from the TDE is still bright.

\begin{figure*}[ht]
\includegraphics[width=\linewidth]{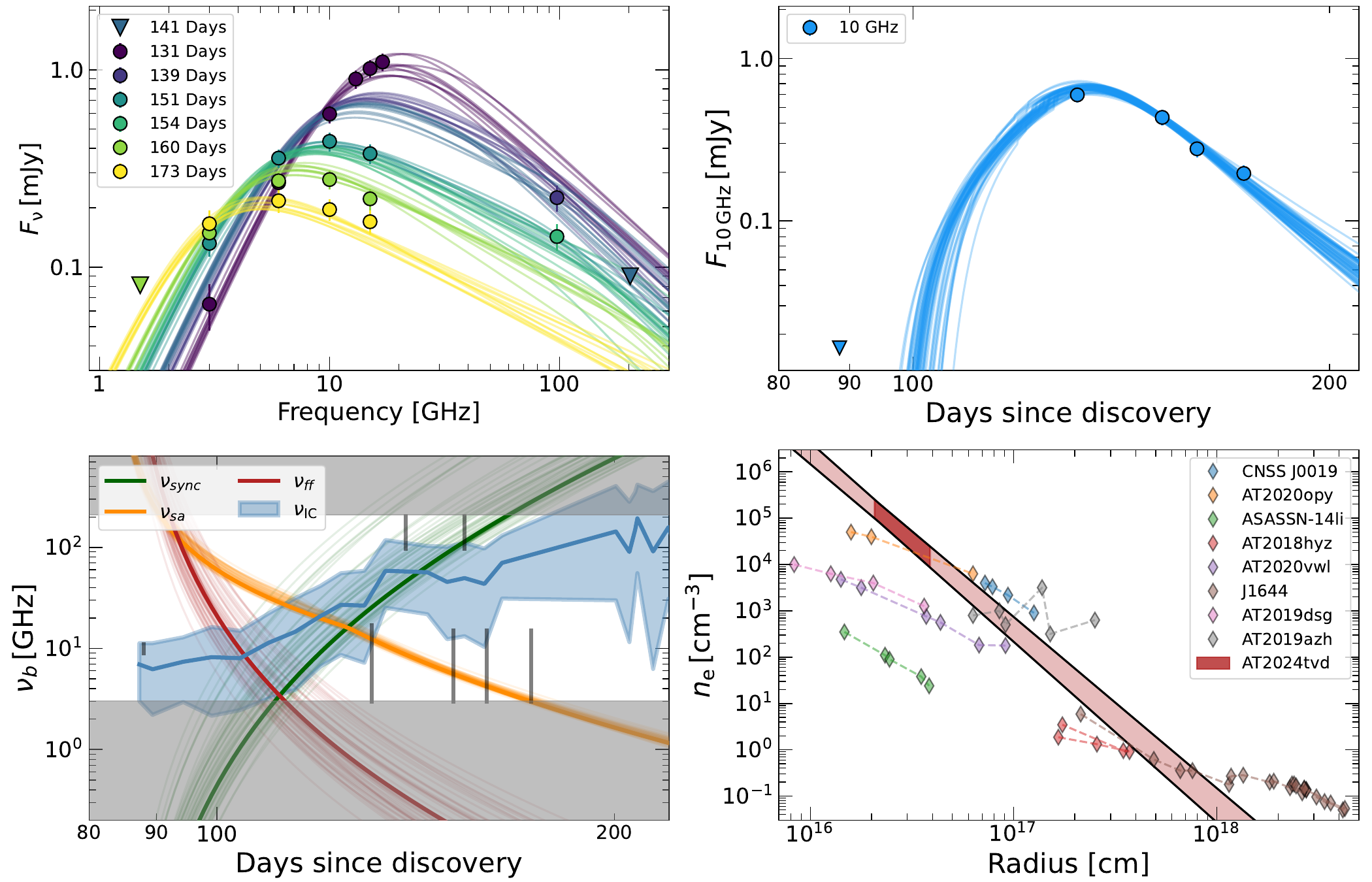}
\caption{Results from the time-dependent model applied to the first radio flare, for $t_{\rm launch} = 84$ days. The \textbf{top left panel} shows the broadband SEDs with models drawn from the posterior distributions obtained by using \texttt{emcee}. The \textbf{top right panel} is for the $10$ GHz light curve with lines drawn from the posterior distributions. The evolution of the broadband SED, and especially the change of the optically thin spectral slope can be easily explained by the transition of the cooling break frequency from $\sim 60$ GHz around $\Delta t = 131$ days to $\gtrsim 100$ GHz at $\Delta t \gtrsim 154$ days. The \textbf{bottom left panel} presents the temporal evolution of the synchrotron break frequencies ($\nu_{\rm m}<0.1$ GHz at all times and is not displayed here). The black vertical lines represent the range of observed frequencies at each time. We note that while we do not fit for $\nu_{\rm IC}$, represented here in a blue region based on the best fitted parameters and the optical/UV luminosity, our observations are consistent, within $1\sigma$ uncertainties with its temporal evolution. Finally, the \textbf{bottom right panel} displays the density profile around AT\,2024tvd compared to other radio bright TDEs (marked with colored diamonds. The density profile we infer for AT\,2024tvd is marked with a red shaded region (marking $1\sigma$ uncertainty), a lighter shade is for the extrapolated radii.
\label{fig: full_ssa_fit}}
\end{figure*}

We fit this model to all of the radio data from the first flare (up to $\Delta t \leq 173$ days) assuming $t_{\rm launch} = 84$ days\footnote{We note that a prompt outflow leads to inconsistencies with the data when IC cooling is accounted for (see Appendix \ref{sec: prompt_sub_rel}). Therefore we assume a delayed outflow and use $t_{\rm launch}$ inferred based on the equipartition analysis in \S\ref{subsec: delayed_outflow}.}. The free parameters are $\beta_{\rm 0}$, $n_{\rm 0}$, $k$, $p$, and $\epsilon_{\rm B}$. In this fitting process we fit for the synchrotron cooling frequency and check that our results are consistent with the location of the IC cooling frequency, $\nu_{\rm IC}$, based on the optical/UV luminosity (Yao et al., in prep.) and the Lorentz factor of the IC cooling (Appendix \ref{sec: prompt_sub_rel}). We do not fit for $\nu_{\rm IC}$ in our fitting process as the uncertainty of the optical/UV luminosity translates to high uncertainties on the fitted parameters. 

We further assume that the temperature of the electrons in the surrounding medium is $T_{\rm e} = 10^5 \, \rm K$, and filling factors of $f_A = 1$, and $f=0.5$. We also assume $\epsilon_{\rm e} = 0.001$ (which is needed for $\epsilon_{\rm B} \leq 0.1$) since there is a degeneracy between $\epsilon_{\rm e}$ and $\epsilon_{\rm B}$. We use \texttt{emcee} to determine the best-fitting parameters and infer their posterior distribution. We use the following flat priors $0.005 < \beta_{\rm 0} < 0.9$, $1 < \log_{10}\left(n_{\rm 0} / \left[ \rm{cm^{-3}} \right]\right) < 11$, $1 < k < 9$, $\log_{10}\left(2.001\right) < \log_{10}\left( p \right) < \log_{10}\left(3\right)$, and $-7 < \log_{10}\left( \epsilon_{\rm B} \right) < \log_{10} \left( 0.33 \right)$. We use $200$ walkers with $30,000$ steps for each chain and discard the first $1,000$ steps for burn-in. We plot the different SEDs, and the $10$ GHz light curve together with lines drawn from the resulting posterior distribution in Fig.~\ref{fig: full_ssa_fit}.

We find that this approach accounts for all of the non-trivial observables of the first radio flare. The early $10$ GHz upper limit is easily explained by the combination of free-free absorption and the delayed outflow. The late emergence of the radio emission and the unprecedentedly fast evolution of the radio spectral peak is explained by a rather steep density profile of $n \left( r \right) = n_{\rm 0} \left( r/r_{\rm 0} \right)^{-3.8 \pm 0.1}$ (see the best-fitting density profile in the bottom right panel of Fig.~\ref{fig: full_ssa_fit}). We also find $n_{\rm 0} = 10^{5.2 \pm 0.2} \rm \, cm^{-3}$ at $r_{\rm 0} \simeq 2.1 \times 10^{16} \, \rm cm$. In addition, the change in the optically thin spectral index between the first and second epochs of our ALMA observations are naturally explained as the result of the synchrotron cooling frequency evolving from $\sim 50$ GHz to $\nu_{\rm c} \gtrsim 100$ GHz at later times. Finally, since our mm-band observations constrain the evolution of the cooling frequency, our fit constrains $\epsilon_{\rm B}$ for the given $\epsilon_{\rm e} = 0.001$. We find $\log_{10} \left( {\epsilon_{\rm B}} \right) = -1.9 \pm 0.3$, implying deviation from equipartition. In the bottom left panel of Fig.~\ref{fig: full_ssa_fit} we plot the evolution of $\nu_{\rm sa}, \nu_{\rm sync}$, and $\nu_{\rm ff}$. The cooling frequency is $\nu_{\rm c} = \min \left[ \nu_{\rm sync}, \nu_{\rm IC} \right]$, and our model is consistent with $\nu_{\rm sync} \simeq \nu_{\rm IC}$ (within $1\sigma$ uncertainty) in all epochs for which we observe a cooling spectral break in the optically thin regime. Therefore, we conclude that our model is self-consistent with the location of the IC cooling frequency even though we are not explicitly accounting for it in our fitting process. On the other hand, we explicitly account for it when we apply the same model for a prompt outflow (which we disfavor because of the required high energetics and densities; see Appendix \ref{sec: prompt_sub_rel}).

\section{Discussion}
\label{sec: discussion}

The radio emission from the first radio-bright off-nuclear TDE\,2024tvd exhibits two rapidly evolving radio flares at late times, with the following main features in order:
\begin{itemize}
    \item An initial deep upper limit, $\nu L_{\rm \nu} \leq 8 \times 10^{36} \, \rm erg \, s^{-1}$, about $88$ days after optical discovery, followed by bright radio emission, $\nu L_{\rm \nu} \sim 3 \times 10^{38} \, \rm erg \, s^{-1}$, at $\Delta t = 131$ days.
    \item A fast decline of the broadband SED, with a peak flux that evolves as $F_{\rm p} \sim t^{-6}$ between $131-173$ days after optical discovery.
    \item The transition of the cooling frequency from $\sim 50$ GHz at $\Delta t = 131-141$ days, to $\gtrsim 100$ GHz at $\Delta t \simeq 154$ days.
    \item Fast re-brightening at $\Delta t = 194$ days, initially rising as $F_{\nu} \sim t^{18}$ in the cm bands, and a fast decay, $F_{\rm \nu} \sim t^{-12}$ of the mm-band emission.
    \item A rise of the peak flux density of the broadband SED in the range of $\Delta t = 218-284$ days, accompanied by a decline of the peak frequency.
\end{itemize}
The radio emission from AT\,2024tvd stands out compared to other radio-bright TDEs. Some TDEs, most notably ASASSN-15oi \citep{Horesh_2021a, Hajela_2025}, and AT\,2020vwl \citep{Goodwin_2023, goodwin_2025}, also exhibit two distinct radio flares. For both ASASSN-15oi and AT\,2020vwl, the separation in time between the two flares is significantly larger than what we observe for AT\,2024tvd, and their radio emission evolves on a much longer timescale (see Fig.~\ref{fig: radio_comparison}). A fast rise is also observed for the late-time component of AT\,2019azh \citep{Sfaradi_2022}. However, it is fainter than both flares of AT\,2024tvd, and it is present in addition to a prompt emission component \citep{Goodwin_2022}. Another notable example is ASASSN-19bt in which a transition in the spectral evolution is observed and associated with either one outflow interacting with a complex CNM, or two distinct outflows \citep{Christy_2024}.

The peculiar evolution of the radio emission from AT\,2024tvd, and specifically the emergence of two late-time radio flares, raises the questions: (1) Can a single outflow produce both flares? (2) When and why were the radio-emitting outflow(s) launched?, and (3) What is the nature of these outflows (i.e., spherical non-relativistic outflow vs. relativistic jet)? We address these questions in the following discussion (a summary of the possible scenarios is reported in Table~\ref{tab: scenarios}).

\begin{table*}[ht]
    \centering
    \begin{tabular}{c|Sc|Sc|}
\cline{2-3}
    & \textbf{First radio flare}
    & \textbf{Second radio flare}\\
\hline
   \pbox{1in}{\textbf{Prompt} \\ \textbf{non-relativistic} \\ \textbf{outflow}}
    & \pbox{2.75in}{Requires high densities ($n_{\rm e} \approx 2 \times 10^{7} \, \rm cm^{-3}$ at $3 \times 10^{16} \, \rm cm$) and post-shock energy (a few $\times 10^{52} \, \rm erg$) to explain the early upper limit and the fast evolution of the broadband SED, therefore it is \underline{\textbf{disfavored}}.}
    & \pbox{2.75in}{The observed fast rise and decay of the radio emission require either a density cavity or larger densities than those inferred for the first radio flare at small radii (producing FFA). We conclude that this scenario is \underline{\textbf{disfavored}}.}\\
\hline
   \pbox{1in}{\textbf{Prompt} \\ \textbf{relativistic} \\ \textbf{outflow}}
    & \pbox{2.75in}{Does not explain the $F_{\rm \nu} \sim t^{-6}$ decay and requires an unphysically steep spectral slope of $p \approx 6$. Therefore, it is unlikely that the first radio flare is a result of a prompt relativistic off-axis jet, and we conclude that this scenario is \underline{\textbf{disfavored}}.}
    & \pbox{2.75in}{The temporal power-laws are extremely fast ($F_{\rm \nu} \sim t^{18}$ during the rise and $F_{\rm \nu} \sim t^{-12}$ in the optically thin regime) and do not follow the closure relations for a prompt off-axis jet. Therefore, this scenario is \underline{\textbf{disfavored}}.}\\
\hline
\pbox{1in}{\textbf{Delayed} \\ \textbf{non-relativistic} \\ \textbf{outflow}}
    & \pbox{2.75in}{An outflow launched at $t_{\rm launch} \approx 84$ days can easily explain the initial upper limit and the fast evolution of the radio emission. This scenario requires a relatively steep density profile, $n_{\rm e} \sim r^{-3.8}$, but with densities similar to other radio-bright TDEs. We find this scenario to be \underline{\textbf{possible}}.}
    & \pbox{2.75in}{There are two possible scenarios of a delayed outflow: (1) the same outflow that produce the first radio flare launched at $t_{\rm launch} \approx 84$ days and interacts with a complex density profile, and (2) a second, mildly-relativistic, delayed outflow launched at $t_{\rm launch} \approx 170$ days. We conclude that both of these scenarios are \underline{\textbf{possible}}.}\\
\hline
\pbox{1in}{\textbf{Delayed} \\ \textbf{relativistic} \\ \textbf{outflow}}
    & \pbox{2.75in}{To achieve a decline of $F_{\rm \nu} \sim t^{-3}$ we require $t_{\rm launch} \sim 80$ days. In this scenario, an observing angle of $\theta \gtrsim 60^{\circ}$ is more likely based on the inferred high densities for lower observing angles. Therefore, we find this scenario to be \underline{\textbf{possible}}.}
    & \pbox{2.75in}{To achieve a decline of $F_{\rm \nu} \sim t^{-3}$ we require $t_{\rm launch} \sim 190$ days. This scenario requires fine-tuning of the jet launching time and the physical parameters for us to observe it and we find it to be \underline{\textbf{possible}} but less likely than the non-relativistic scenario.}\\
\hline
\end{tabular}
    \caption{A summary of the different possible scenarios for each radio flare.\label{tab: scenarios}}
\end{table*}

\subsection{The nature of the first radio flare}
\label{subsec: first_flare_discussion}

Newtonian equipartition analysis of the emission from the first radio flare suggests that for an outflow that was launched promptly after the optical discovery, the shock is accelerating. The shock velocities we infer in this case are $\sim 0.06-0.085c$, and are in agreement with the velocities expected from accretion-driven winds and the unbound tidal debris stream \citep{Strubbe_2009, Krolik_2016, bonnerot_2020, alexander_2020}. However, a prompt outflow from AT\,2024tvd requires an extremely high post-shock energy, of a few $\times 10^{52} \, \rm erg$, and CNM densities, $n_{\rm e} \sim 2 \times 10^7 \, \rm cm^{-3}$ at $\sim 3 \times 10^{16} \, \rm cm$ (see Appendix \ref{sec: prompt_sub_rel}). Instead, our time-dependent analysis suggests that a delayed Newtonian outflow, launched at $t_{\rm launch} = 84^{+6} _{-25}$ days, naturally explains all emission features without invoking extreme energetics and densities, thus favoring the delayed outflow scenario in the context of non-relativistic ejecta.

Physically, a delayed emergence of the radio emission might arise from a relativistic jet that initially points away from our line of sight (e.g., \citealt{Matsumoto_2023, Sfaradi_2024} in the context of TDEs). However, in this scenario it is challenging to explain the fast decay of $F_{\rm \nu} \sim t^{-6}$ with the expected $F_{\rm \nu} \sim t^{-p}$ \citep{Beniamini_2023_J1644}, which would imply a nonphysical spectral slope of $p \approx 6$. Therefore, we find that a prompt off-axis jet is an unlikely scenario. We estimate that the earliest outflow launch time for an off-axis jet is at $\Delta t \approx 80$ days (by requiring $F_{\rm \nu} \sim t^{-3}$), similar to the inferred $t_{\rm launch}$ for a non-relativistic outflow. In this context, we favor $\theta \gtrsim 60^{\circ}$ as smaller viewing angles require higher Lorentz factors and higher densities, which are less physical at the inferred large radii (see top panel of Fig.~\ref{fig: off_axis_equipartition}).

Overall, our analysis points at a delayed outflow (either non-relativistic or ultra-relativistic), launched at $\Delta t \sim 80$ days for the origin of the first radio flare. Interestingly, the X-ray analysis conducted by \cite{Yao_2025} concluded that the X-ray emission during the first $\sim 80$ days is best described by a thermal multi-temperature disk model with no evidence for a Comptonization component. A Comptonization component emerges at $\Delta t \gtrsim 80$ days (see Fig. 5 in \citealt{Yao_2025}). We find it intriguing that the onset of a detectable Comptonization in the X-rays is roughly simultaneous with the inferred radio outflow launching time, and speculate that the origin of the first radio flare is an accretion-driven outflow. Thermal X-ray emission from TDEs is typically associated with the accretion around the MBH. Therefore, it is possible that the spectral change in the X-ray emission is related to the accretion state of the disk. A disk-outflow relation is used to explain the observed radio--X-ray connection for X-ray binaries \citep[XRBs;][]{Fender_2004}. There, the emergence of optically thin radio emission is associated with a jet launched following a transition in the accretion state seen in the X-rays. In some cases discussed in the context of XRBs, the Comptonization component in the hard X-rays can be a result of up-scattering of thermal disk photons by a hot $e^{-}$ corona and/or dynamical Comptonization by in-falling material (e.g., \citealt{titarchuk_2021}). If the emergence of the non-thermal X-ray component is due to the latter, a fraction of the in-falling material can result in an outflow. Finally, a disk-outflow connection was also proposed for TDE accretion disks at late times when the accretion rate decreases below $\sim 0.03 \dot{M}_{\rm Edd}$ \citep{Giannios_2011}. However, the inferred outflow launch day ($\Delta t \sim 80$ days) is probably too early for the accretion rate to drop to these sub-Eddington values.

\subsection{The nature of the second radio flare}
\label{subsec: second_flare_discussion}

The origin of the second radio flare is difficult to explain with a prompt outflow scenario (both non-relativistic and relativistic), as it was first detected at $\Delta t \approx 194$ days, and is initially rising as fast as $F_{\rm \nu} \sim t^{18}$ (with a transition to $F_{\rm \nu} \sim t^{4}$ about a month later). A prompt non-relativistic outflow would require either large densities at $R\leq 6 \times 10^{16} \rm \, cm$ to explain the late turn on of the radio emission with absorption in front of the shock suppressing the radio emission or a density cavity at these radii. In addition, a density enhancement at $R \geq 6 \times 10^{16} \, \rm cm$ is needed if both radio flares are from the same non-relativistic outflow (in the equipartition formalism). Therefore, the second radio flare is less likely to be a result of a prompt non-relativistic outflow. On the other hand, the observed increase in the equipartition energy, and the fast evolution of the optically thin emission can also be explained by an off-axis relativistic jet, as emission from more parts of the jet are entering our line of sight at late-times (see e.g., Swift J1644+57; \citealt{Beniamini_2023_J1644}, AT\,2018hyz; \citealt{Sfaradi_2024}). However, we find that even for an off-axis jet $t_{\rm launch} \simeq 190$ days is needed to match $F_{\rm \nu} \sim t^{-3}$ in the optically thin regime. This is only $\sim 4$ days before the first radio detection of this flare. Therefore we conclude that in the off-axis jet scenario, the jet is either only mildly-relativistic if it is significantly off-axis ($\Gamma \beta \lesssim 1$ for $\theta \approx 90^{\circ}$), or, if the jet is ultra-relativistic, it is only slightly off-axis ($\Gamma \simeq 12$ for $\theta \approx 30^{\circ}$). Although we find this scenario possible, we note that it requires some fine-tuning of the physical parameters.

Analysis of the broadband SEDs of the second flare in the Newtonian limit results in a radio emitting outflow at larger radii than the radii inferred for the first flare (for a non-relativistic outflow). In addition, the first estimation of the equipartition radius during this second flare is consistent with the temporal evolution of the equipartition radius during the first radio flare. Therefore, it is possible that a single outflow is producing both radio flares. The inferred density profile in a single outflow scenario, assuming $t_{\rm launch} = 84^{+6} _{-25}$ days (which is the preferred outflow launch time), points to a flattening of the density profile followed by a shallower density profile. For this reason, we consider the possibility that the rebrightening of the radio emission is due to one delayed-outflow interacting with a complex density profile (as was previously suggested by \citealt{Matsumoto_2024}). Alternatively, it is also possible that a second outflow was launched at $t_{\rm launch} = 170 \pm 10$ days. A higher shock velocity of $\sim 0.5c$ is needed for such an outflow, and lower densities, roughly matching the extrapolation of the first delayed radio flare in the framework of the equipartition analysis, but with a shallower profile of $n_{\rm e} \sim r^{-2}$. 

Overall, our analysis of the emission from the second flare favors either (1) a scenario in which the same outflow launched around $t_{\rm launch} \approx 80$ days and interacting with a complex density profile or (2) a scenario of a second distinct outflow launched at $t_{\rm launch} \approx 170$ days for a sub-relativistic outflow or $t_{\rm launch} \approx 190$ days for a relativistic outflow. We note that the high-GHz emission observed with ALMA at $\Delta t = 227$ days is too high compared to the emission in low-GHz, and we cannot account for it in our equipartition analysis. We leave the analysis of this emission to future time-dependent modeling. While observations are still being acquired and we defer a full time-dependent modeling of the second flare to later paper, we discuss the astrophysical implications of our findings below.

\subsection{Astrophysical implications}
\label{subsec: two_flares_discussion}

AT\,2024tvd is the first bonafide off-nuclear TDEs with extensive radio coverage and it also exhibits one of the most peculiar radio light curves observed for a TDE to date. We raise here the possibility that the unique radio evolution is related to the off-nuclear position of the MBH and to the possible IMBH-TDE scenario (\citealt{Yao_2025} inferred that the black hole mass can be as low as $10^{5} \, \rm M_{\odot}$). For example, if the two flares are a result of multiple accretion-driven outflows, or discrete ejections of a dense blobs of material (also referred to as a "knot ejection"; see e.g., \citealt{king_2016}), it is unclear why multiple outflows are not observed in all TDEs, and can be related to the different mass-scaling if AT\,2024tvd is an IMBH-TDE. Furthermore, if the double-peaked radio light-curve is a result of a complex density profile it can be related to the origin of this off-nuclear MBH (i.e., a recoiling MBH or an MBH from a minor galaxy merger) and the stellar population around it. On the other hand, it is important to note that there is a wide range of radio properties observed for different TDEs. Therefore, it is possible that this unique behavior is just another layer of an already complex landscape of TDEs in the radio. In the following discussion, we explore several astrophysical scenarios for the radio emission from AT\,2024tvd in an attempt to explain the origin of both radio flares with a single outflow.

\cite{Matsumoto_2024} suggested that double-peaked radio light curves can result from the flattening of the density profile at the Bondi radius. Interestingly, in our equipartition analysis we find that the outflow radius at the time of the second radio flare is $\sim 6 \times 10^{16} \, \rm cm$, similar to the Bondi radius inferred by \cite{Matsumoto_2024} for a $10^{6} \, M_{\odot}$ black hole and an ISM temperature of $10^{7} \, \rm K$. However, this temperature is higher than expected for a warm ISM. On the other hand, as the environment of a non-AGN MBH is determined by the mass and energy injection from the surrounding stellar population \citep{quataert_2004, generozov_2015}, a density break is also expected at the stagnation radius, $R_{\rm s}$, that separates the inflow to the MBH from the outwards stellar winds \citep{generozov_2015, yalinewich_2018}:
\begin{align}
\label{eq: stag_radius}
    R_{\rm s} \simeq & 2.5 \frac{G M_{\rm BH}} {v_{\rm w}^2} \\
    \nonumber \simeq & 3 \times 10^{16} \, {\rm cm} \, \left( \frac{M_{\rm BH}}{10^{6} \, M_{\rm \odot}} \right) \left( \frac{v_{\rm w}}{1000 \, {\rm km \, s^{-1}}} \right)^{-2},
\end{align}
where $M_{\rm BH}$ is the mass of the black hole, and $v_{\rm w}$ is related to the heating of the gas by stellar winds. Based on the normalization of Eq.~\ref{eq: stag_radius}, and the density we infer at $\approx R_{\rm s}$, a high pre-TDE accretion rate is expected for a $10^{6} \, M_{\odot}$ black hole:
\begin{align}
    \dot{M} \simeq & 4 \pi R_{\rm s}^2 m_{\rm p} n_{\rm e} \left( R_{\rm s} \right) v_{\rm ff} \\
    \nonumber \simeq & 0.03 \dot{M}_{\rm Edd} \left( \frac{n_{\rm e} \left( R_{\rm s}\right)}{4 \times 10^{4} \, {\rm cm^{-3}}} \right) \left( \frac{R_{\rm s}}{3 \times 10^{16} \, {\rm cm}} \right)^{3/2} \\
    \nonumber & \times \left( \frac{M_{\rm BH}}{10^{6} \, M_{\odot}}\right)^{-1/2},
\end{align}
where the free-fall velocity $v_{\rm ff} \equiv \left(\frac{G M_{\rm BH}}{R_{\rm s}}\right)^{1/2}$ is estimated at the stagnation radius. Therefore, if the double-peaked radio light curve is a result of a break in the density profile at the stagnation radius, high accretion rate and wind velocities are needed. This implies a very young stellar population of $\tau_{*} \sim 10^7$ years around this off-nuclear MBH \citep{generozov_2015}. Eq.~\ref{eq: stag_radius} is scaled to achieve a stagnation radius roughly at the same radius that we infer for the density break for $M_{\rm BH} \gtrsim 10^{6} \, \rm M_{\odot}$. In order to achieve the same radius in the IMBH scenario, lower winds velocities are required, which result in an older stellar population. While the presence of the stagnation radius explains the observed break in the density profile, the slopes of the inferred density profiles, especially at $r<R_{\rm s}$, are relatively steep and are not easily explained in this scenario. 

A different natural explanation for the break in the density profile is that the radio-emitting shock first interacts with either the stellar debris or an early launched spherical outflow, and only later interacts with the CNM. Since these early outflows are expected to travel at a velocity of a few $\times 10^{4} \, \rm km \, s^{-1}$, if they are launched around the time of optical discovery, they can reach a radius of a few $\times 10^{16} \, \rm cm$ at the time of first radio detection ($\Delta t = 131$ days). Therefore, it is possible that an accretion-driven outflow that was launched at $\Delta t \approx 80$ days after optical discovery is first catching up with this early outflow, producing the first radio flare, and that the late re-brightening is due to the transition of the outflow to the CNM. This scenario can potentially explain the two distinct density structures and the steep density profile as being measured at the edge of an early spherical outflow or of the stellar debris. On the other hand, our analysis assumes a spherical outflow and the geometry of the stellar debris in this scenario is inconsistent with the radii we infer. That is, the stellar debris stream is expected to be gravitationally bound and stretch only a few tens of solar radii in diameter \citep{Coughlin_2016}. For a gravitationally bound stellar stream stretching a few $\times 10^{16} \, \rm cm$, and with a width of $\sim 10^{12} \rm \, cm$, the radio-emitting region appears smaller than it appears for a spherical interaction region and has an area filling factor of $f_{\rm A} \sim 10^{-4}$. The radius of the emitting region scales as $f_{\rm A}^{-8/19}$ \citep{Krolik_2016}. Therefore, the radii we infer in the case of an interaction with a gravitationally bound stream are larger by more than an order of magnitude than for a spherical outflow and are not consistent with a sub-relativistic tidal debris stream. We also emphasize that while the equipartition radius and the double-peaked radio light curve motivate a broken power-law density profile around the MBH, we cannot explain the accelerating shock during the second radio flare.

The discussion above is focused on a sub-relativistic outflow. However, a double-peaked radio light curve is also expected, in some cases, for relativistic jets observed off-axis. For example, a double-peaked non-thermal emission component is sometimes discussed in GRBs for some viewing angles if the jet is structured \citep{beniamini_2020_structured_jet}, or, when the reverse shock crosses the ejecta as it decelerates, resulting in two emission zones peaking at different times \citep{Abdikamalov_2025}. However, in the case of AT\,2024tvd, even if we set the jet launch day to $t_{\rm launch} \approx 80$ days the temporal evolution of the optically thin emission during the rebrightening is $F_{\rm \nu} \sim t^{-8}$, which is too fast even for an off-axis jet (see Fig.~\ref{fig: light_curves_dalayed}). \cite{Teboul_2023} presented another scenario in which some relativistic jets launched by TDEs can only be observed at late times (see also \citealt{lu_2024}). In cases of misalignment between the spin of the MBH and the orbital plane of the disrupted star, the disk/jet structure will precess and a large quasi-spherical structure might be formed. The jet can then escape prior to, or at, the alignment with the spin of the MBH, depending on the alignment mechanism, i.e., magneto-spin vs. hydrodynamic. Based on the observed timescales for AT\,2024tvd, and for an MBH with $10^{6} \, \rm M_{\odot}$, hydrodynamic alignment of the disk/jet system, due to internal stresses within the accretion disk is more likely. Furthermore, \cite{Teboul_2023} find that for such black hole mass, and for a disk viscosity parameter $\alpha \simeq 0.1$, the alignment timescale is $\sim 193$ days, similar to our first detection of the second radio component. Therefore, we consider it plausible that both radio flares are a result of the same jet entering our line of sight as it precesses, and that the onset of the second radio flare is at the alignment of the orbital plane with the MBH spin. However, in that case, the jet needs to align directly into our line of sight, or significantly decelerate and expand to a quasi-spherical outflow. This is somewhat supported by the high velocities we infer during the second radio flare even in the Newtonian regime. Further modeling is needed to determine the likelihood of this scenario.

\section{Summary and conclusions}
\label{sec:conclusions}

This work presents the radio emission from the first optically selected off-nuclear TDE, AT\,2024tvd. Our extensive radio campaign results in broadband observations in the range of 1.5--230 GHz, spanning from $88$ days to $\sim 10$ months after optical discovery, making AT\,2024tvd one of the best studied TDEs in radio wavelengths so far. In addition to its non-traditional position $\sim 0.8$ kpc from the host nucleus, the temporal evolution of the radio emission from this TDE is unprecedented. Its double-peaked radio light curve is evolving faster than any other known TDE, with $F_{\rm \nu} \sim t^{-6}$ and $\sim t^{-12}$ during the first and second radio components, respectively. In addition, the broadband SEDs revealed changes in the spectral slope of the optically thin regime and a monotonically declining radio spectral peak during the first radio component. However, a fast optically thin decline and an increasing radio spectral peak are observed during the second radio component.

Our analysis focuses on the nature of the radio emitting outflow/outflows, i.e., relativistic vs. non-relativistic, and their launch time (promptly after optical discovery vs. delayed outflow). The overall emission is consistent with either non-relativistic, mildly-relativistic, and/or ultra-relativistic outflows. However, based on the temporal indices, it is unlikely that an outflow that was launched at, or around, the time of optical discovery is responsible for any of the radio components, and there is strong evidence for delayed outflow(s). The first radio component is consistent with an outflow that is launched at $t_{\rm launch} \approx 80$ days, which coincides with the emergence of comptonized spectral component in the X-rays, possibly hinting toward an accretion-driven outflow. 

The second flare is consistent with either a mildly-relativistic, $\sim 0.5c$, outflow launched at $t_{\rm launch} \approx 170$ days, or a relativistic jet (up to $\Gamma \sim 12$ for $\theta = 30^{\circ}$, launched at $t_{\rm launch} \approx 190$ days; only $\sim 4$ days before the first detection of the second radio component). On the other hand, we note that we cannot rule out a scenario in which both flares are from the same sub-relativistic outflow that was launched at $t_{\rm launch} \approx 80$ days. In this case the double-peaked radio light curve is a manifestation of a shock wave interacting with a broken power-law density profile. We show that such a density break is expected around the stagnation radius. We also raise the possibility that either the stellar debris or an early launched spherical outflow are the medium with which the delayed outflow first interacts before reaching the more distant CNM, and hence the possible transition in the density profile.

It is also worth highlighting that AT\,2024tvd is the first radio bright, bonafide off-nuclear TDE, and it is also the TDE with the fastest evolution observed to date. We therefore put forward the idea of a connection between the unique radio emission and the off-nuclear (and possibly intermediate mass) origin of the massive black hole. On the other hand, it is also relevant to point out that the evolution of the thermal emission from this TDE observed in optical--UV--X-ray wavelengths is rather standard for a TDE \citep{Yao_2025}. Furthermore, radio emission from TDEs does not seem to follow a well-defined evolution, and the emission has been observed to evolve quickly and peak on different timescales for different TDEs. Therefore, it is also possible that the unique evolution of the radio emission and the off-nuclear position are not related. 

In order to break some of the model degeneracies and determine if the radio emitting outflows are relativistic or non-relativistic, high-resolution, Very Long Baseline Interferometry (VLBI) observations are required to resolve a jet structure and/or detect the movement of a relativistic jet. Polarization measurements at different times can also help differentiate between sub-relativistic and relativistic outflows since the emission is expected to be more polarized around the time of the peak of a jetted outflow \citep{gill_2018}. In addition, early observations in the mm-bands are sometimes the only way to constrain the outflow launch date as an absorbed emission is expected to first peak in those bands. We plan to continue monitoring the radio emission from AT\,2024tvd in an attempt to better understand the nature of this event.

\section*{Acknowledgments} \label{sec:acknowledgments}

The National Radio Astronomy Observatory (NRAO) is a facility of the National Science Foundation operated under cooperative agreement by Associated Universities, Inc. We thank the NRAO for carrying out the Karl G. Jansky Very Large Array (VLA) and the Atacama Large Millimeter Array (ALMA) observations. 
This paper makes use of the following ALMA data: ADS/JAO.ALMA\#2024.A.00024.T, and ADS/JAO.ALMA\#2024.A.00034.T. ALMA
is a partnership of ESO (representing its member states), NSF (USA) and NINS (Japan),
together with NRC (Canada), MOST and ASIAA (Taiwan), and KASI (Republic of Korea), in
cooperation with the Republic of Chile. The Joint ALMA Observatory is operated by
ESO, AUI/NRAO and NAOJ. 
The Allen Telescope Array refurbishment program and its ongoing operations are being substantially funded through the Franklin Antonio Bequest. Additional contributions from Frank Levinson, Greg Papadopoulos, the Breakthrough Listen Initiative and other private donors have been instrumental in the renewal of the ATA. Breakthrough Listen is managed by the Breakthrough Initiatives, sponsored by the Breakthrough Prize Foundation. The Paul G. Allen Family Foundation provided major support for the design and construction of the ATA, alongside contributions from Nathan Myhrvold, Xilinx Corporation, Sun Microsystems, and other private donors. The ATA has also been supported by contributions from the US Naval Observatory and the US National Science Foundation. 
We acknowledge the staff who operate and run the AMI-LA telescope at Lord's Bridge, Cambridge, for the AMI-LA radio data. AMI is supported by the Universities of Cambridge and Oxford, and by the European Research Council under grant ERC-2012-StG-307215 LODESTONE. 
The Submillimeter Array is a joint project between the Smithsonian Astrophysical Observatory and the Academia Sinica Institute of Astronomy and Astrophysics and is funded by the Smithsonian Institution and the Academia Sinica.  We recognize that Maunakea is a culturally important site for the indigenous Hawaiian people; we are privileged to study the cosmos from its summit.

R.~M. acknowledges partial support from the National Science Foundation (grant number AST-2224255).
KDA and CTC acknowledge support provided by the NSF through award SOSPA9-007 from the NRAO and award AST-2307668. KDA gratefully acknowledges support from the Alfred P. Sloan Foundation.
B.~D.~M. acknowledges partial support from the National Science Foundation (grant number AST-2406637) and the Simons Foundation (grant number 727700). The Flatiron Institute is supported by the Simons Foundation.
P.B. was supported by a grant (no. 2020747) from the United States-Israel Binational Science Foundation (BSF), Jerusalem, Israel (PB) and by a grant (no. 1649/23) from the Israel Science Foundation
AJG is grateful for support from the Forrest Research Foundation.
N.F. acknowledges support from the National Science Foundation Graduate Research Fellowship Program under Grant No. DGE-2137419.
R.B.D. acknowledges support from the National Science Foundation under grant 2107932.
This work was supported by the Australian government through the Australian Research Council’s Discovery Projects funding scheme (DP200102471).

\begin{acknowledgments}
\end{acknowledgments}

\software{astropy \citep{2013A&A...558A..33A,2018AJ....156..123A},  
\texttt{emcee} \citep{foreman_mackey_2013},
CASA \citep{CASA}
}

\appendix

\section{Prompt sub-relativistic outflow fit}
\label{sec: prompt_sub_rel}

In \S~\ref{subsec: multi_epoch} we considered a delayed, sub-relativistic, outflow as the origin of the first flare. We showed that our multi-epoch fit for such an outflow explains all the observed properties of the first flare. We also noted that the same fitting procedure requires extreme outflow parameters under the assumption of a prompt sub-relativistic. Here, we briefly present the results of the prompt outflow fitting process, discuss its implications, and conclude that it is disfavored.

We consider the same model discussed in \S~\ref{subsec: multi_epoch}, but we introduce three main changes: (i) we adopt $t_{\rm launch} = 0$ days, (ii) $\epsilon_{\rm e}$ is a free parameter, this is now possible since this model requires high densities to explain the first upper limit and we are provided with a measurement of the mass in front of the shock by the FFA, therefore, $\epsilon_{\rm e}$ and $\epsilon_{\rm B}$ are no longer degenerate, and (iii) we define $\nu_{\rm c} = \min \left[ \nu_{\rm sync}, \nu_{\rm IC} \right]$. To calculate $\nu_{\rm IC}$ we use the IC cooling Lorentz factor (e.g., \citealt{Nayana_2024})
\begin{align}
    \label{eq: gamma_IC}
    \gamma_{\rm IC} = 1.16 \times 10^{19} \frac{r^2}{\Gamma \left(\Delta t - t_{\rm launch} \right) L_{\rm bol}},
\end{align}
where $r$ is the radius of the radio emitting shock, $\Gamma$ is the bulk Lorentz factor, and we use the thermal optical/UV luminosity during the time of the first radio flare $L_{\rm bol} = 4.5 \times 10^{43} \left(\frac{\Delta t}{68 \, \rm{days}}\right)^{-1.6} \, \rm erg \, s^{-1}$ (Yao et al., in prep.). We do not consider here the uncertainty on $L_{\rm bol}$ but note that it has negligible impact on our conclusions.

\begin{figure}[ht]
\centering
\includegraphics[width=0.49\linewidth]{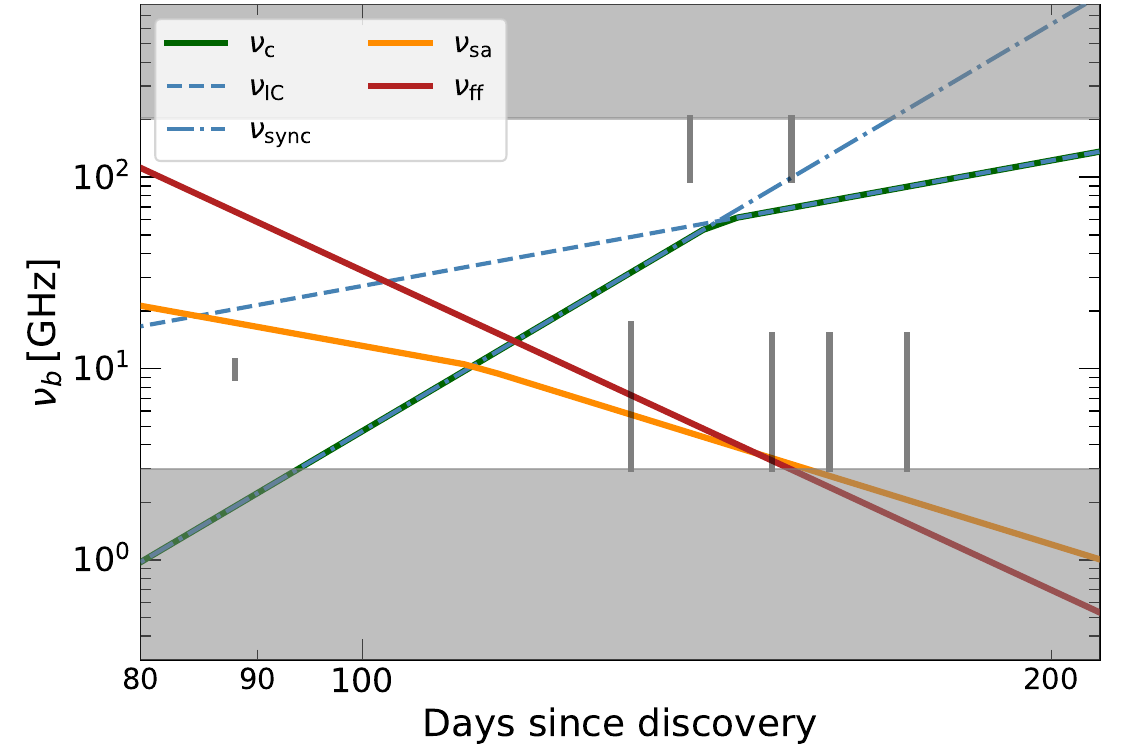}
\\
\includegraphics[width=0.49\linewidth]{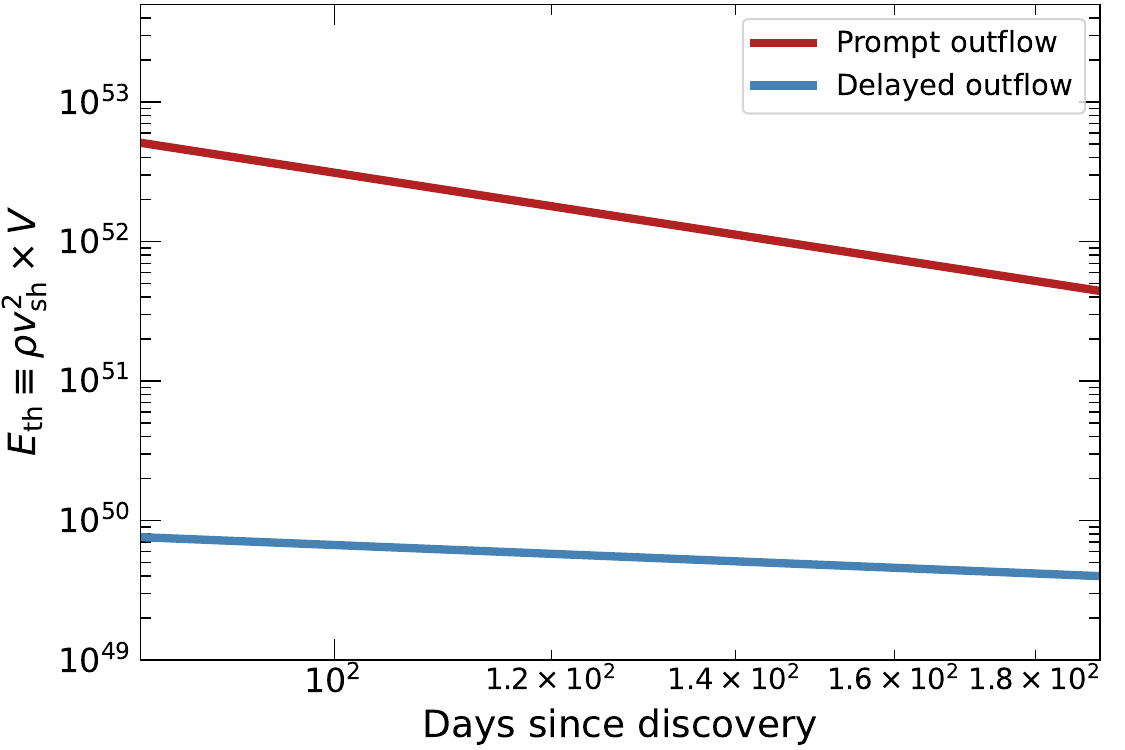}
\includegraphics[width=0.49\linewidth]{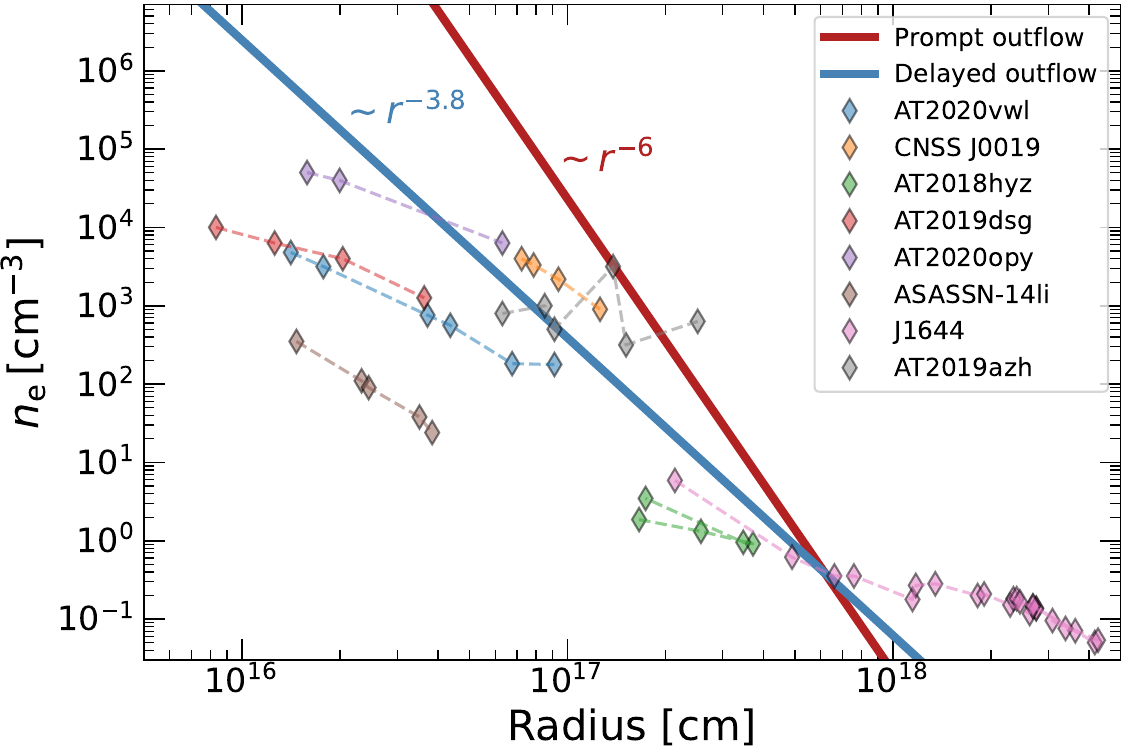}
\caption{Results from our time-dependent fit of the first radio flare assuming a prompt non-relativistic outflow. The top panel is for the temporal evolution of the different break frequencies based on our best-fitting parameters: $\beta_{\rm 0} = 0.14 \pm 0.02$, $\log \left( n_{\rm 0} \, {\rm \left[ cm^{-3}\right]}\right) = 7.36 \pm 0.07$, $k=6.05 \pm 0.25$, $p=2.05 \pm 0.05$, $\log \left( \epsilon_{\rm e}\right) = -4.6 \pm 0.5$, and, $\log \left( \epsilon_{\rm B}\right) = -3.5 \pm 0.2$. The bottom left panel shows the post-shock energy for both the prompt and delayed outflows. Finally, the bottom right panel presents the density profiles we infer in each scenario compared to other radio-bright TDEs. Both the energetics and densities inferred in the prompt outflow scenario are extreme when compared to the other scenarios we explored in this work and other radio-bright TDEs.
\label{fig: multi_epoch_prompt_vs_delayed}}
\end{figure}

We present in Fig.~\ref{fig: multi_epoch_prompt_vs_delayed} the evolution of the break frequencies, the inferred energy, and the density profiles of our delayed and prompt non-relativistic outflow models. We find that while there is a set of parameters that provide a statistically acceptable fit to the observed broadband radio emission (including the early upper limit), the inferred values are physically extreme. For example, the post-shock energy is few$\times 10^{52} \, \rm erg$, significantly higher than the energy we infer for any other scenario and for other radio-bright TDEs with sub-relativistic outflows. Furthermore, the density profile is extremely steep, $n_{\rm e} \sim r^{-6}$, and implies higher densities at ($R \lesssim 10^{17} \, \rm cm$) than any other radio-bright TDE. Based on this analysis, we conclude that the origin of the first flare is unlikely to be related to a prompt, non-relativistic outflow.

\section{Emission from the host galaxy nucleus}
\label{sec: host_emission}

\begin{figure}[ht]
\centering
\includegraphics[width=0.5\linewidth]{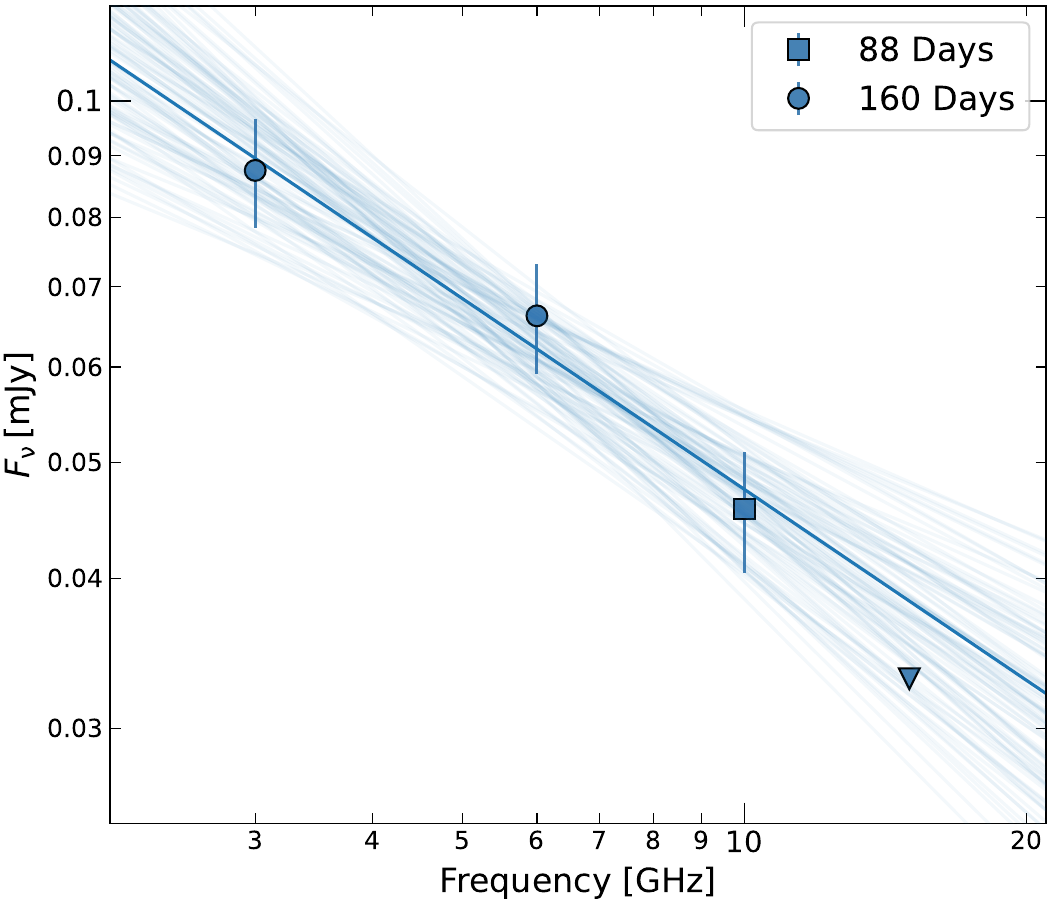}
\caption{Broad-band SED of the emission from the host-galaxy nucleus. The filled circles are for radio detections obtained at $\Delta t = 160$ days, the square is for the detection obtained at $\Delta t = 88$ days, and the triangle is a $3\sigma$ upper limit obtained at $\Delta t = 160$ days. Also plotted is the best-fitting power-law (solid line) and models drawn from the posterior distribution.
\label{fig: host_emission}}
\end{figure}

As mentioned in \S\ref{sec:radio_observations}, some of our observations showed two point sources: one at the known optical position of the TDE\,2024tvd, and the other at the optical position of the center of the host galaxy (see coordinates reported in \citealt{Yao_2025}). We construct an SED of the host galaxy nucleus by taking flux measurements at two different epochs from VLA observations taken in A-configuration: $\Delta t = 88$ days for the X-band image, and $\Delta t = 160$ days for the S- C- and Ku-bands (Fig.~\ref{fig: host_emission}). The emission is optically thin, and we find a best-fitting $F_{\rm \nu} = A \left( \frac{\nu}{5 \, \rm GHz} \right)^{-\beta}$ with $A = 64 \pm 4\, \rm \mu Jy$, and $\beta = 0.53 \pm 0.12$. Based on this fit, we estimate $L_{\rm 1.4 \, GHz} = \left( 3.2 \pm 0.5 \right) \times 10^{37} \, \rm erg \, s^{-1} Hz^{-1}$. Adopting the relation between the star formation rate (SFR) and the $1.4$ GHz luminosity from \cite{davies_2017}, we infer an SFR of $0.51 \pm 0.06 \, \rm M_{\rm \odot} \, yr^{-1}$. This value is significantly higher than the SFR constraints, $\leq 0.1 \, \rm M_{\odot} \, yr^{-1}$, derived from stellar population synthesis analysis \citep{maraston_2009}. Therefore, as concluded by \cite{Yao_2025}, the radio source at the center of the galaxy must be powered by a low-luminosity AGN with a steep radio spectrum, possibly associated with a jet \citep{merloni_2003}.

We use this host-galaxy radio model to estimate the contamination component for the observations obtained with the VLA when it was in C- and D-configurations and subtract it from the TDE emission (in S-, C-, and X-bands). However, we note that this is a minor correction of less than $10\%$ of the flux measurements for these observations.

\section{Synchrotron self-absorption frequency in the slow-cooling regime}
\label{sec: ssa_frequency}

The synchrotron emission from astrophysical transients has been studied extensively and multiple frameworks have been developed to infer the physical parameters from observed radio SEDs (e.g., \citealt{Sari_1998, chevalier_1998,granot_2002,BDNP13,matsumoto_2021,Matsumoto_2023,Margalit_2021}). Sometimes assumptions from non-equipartition frameworks have been combined in the TDE literature, leading to results that are not self-consistent. Thus motivated, in the next section we derive all the synchrotron break frequencies for different ordering of $\nu_{\rm m}, \nu_{\rm sa}, \nu_{\rm c}$ in the \emph{slow cooling} regime ($\nu_{\rm m} < \nu_{\rm c}$; \citealt{Sari_1998}), for a non-relativistic outflow, and provide a recipe for calculating synchrotron SEDs (we provide the code used to calculate the synchrotron SEDs based on this method at \url{https://github.com/Itaisfa/full_synchrotron_SED/tree/main}). 

Consider a shock moving with a bulk Lorentz factor $\Gamma$, a velocity $v_{\rm sh}$, and located at a redshift $z$ and a luminosity distance $d_{\rm l}$. At the shock front electrons are accelerated to a power-law energy distribution of the form:
\begin{align}
\label{eq: energy_distribution}
    N\left(\gamma\right) d \gamma = N_{\rm 0} \left( \gamma/ \gamma_m \right)^{-p} d \gamma; \, \gamma > \gamma_{\rm m}
\end{align}
with
\begin{align}
    \gamma_{\rm m} = {\max} \left[1, 4 \epsilon_{\rm e} \frac{p-2}{p-1} \frac{m_{\rm p}}{m_{\rm e}} \left( \frac{v_{\rm sh}}{c} \right)^2 \right]
\end{align} 
where the factor of $4$ arise from the definition of $\epsilon_{\rm e}$ as a fraction of post shock energy that goes to the relativistic electrons and the Rankine–Hugoniot jump conditions; $m_{\rm e}$ and $m_{\rm p}$ are the electron and proton masses, respectively. \cite{matsumoto_2021} considered a minimum Lorentz factor of $2$. Here we used a minimum $\gamma_{\rm m} = 1$ for self-consistency with \cite{chevalier_1998}. We note that in the ultra-relativistic regime the factor of $\left( v_{\rm sh}/c\right)^2$ should be replaced with $\left(\Gamma -1\right)$ \citep{BDNP13, Beniamini_2023_J1644}. We also find, for $p>2$, that $N_{\rm 0} \equiv N \left( \gamma = \gamma_{\rm m} \right)$ is:
\begin{align}
    N_{\rm 0} = \frac{\epsilon_{\rm e}}{\epsilon_{\rm B}} \frac{B^2 \left( p-2 \right)}{8 \pi \gamma_{\rm m}^2 m_{\rm e} c^2}
\end{align} by calculating $E_{\rm e} = \int^{\infty} _{\gamma_{\rm m}} N_{\rm 0} \left(\gamma/\gamma_{\rm m}\right)^{-p}\left( \gamma m_{\rm e} c^2 \right)d\gamma$ and requiring $\frac{E_{\rm e}}{\epsilon_{\rm e}} = \frac{E_{\rm B}}{\epsilon_{\rm B}}$. The critical Lorentz factor below which the electron does not lose a significant fraction of its energy to radiation (i.e., the synchrotron cooling Lorentz factor) is:
\begin{align}
    \gamma_{\rm c} = \frac{6 \pi m_{\rm e} c}{\sigma_{\rm T} \Gamma B^{2} t}
\end{align}
where $B$ is the magnetic field strength, $\sigma_{\rm T}$ is the Thomson cross-section, and $c$ is the speed of light. The characteristic synchrotron frequency of an electron with Lorentz factor $\gamma$ is
\begin{align}
\label{eq: synchrotron_char_freq}
    \nu \left( \gamma \right) = \frac{q_{\rm e} B \Gamma \gamma^2}{2 \pi m_{\rm e}c \left( 1+z\right)}
\end{align}
where $q_{\rm e}$ is the electric charge. Now we can derive the minimal and cooling frequencies, $\nu_{\rm m}$ and $\nu_{\rm c}$, by replacing $\gamma_{\rm m}$ and $\gamma_{\rm c}$, respectively, with $\gamma$ in Eq.~\ref{eq: synchrotron_char_freq}.

Following \cite{Sari_1998}, the flux density at the peak of the synchrotron emission is\footnote{As mentioned in \cite{BDNP13}, the precise numerical pre-factor depends on the power-law index of the energy distribution of the electrons - $p$. We provide the numerical pre-factor considered by taking the peak spectral power $P_{\rm \nu, \, max}$ and multiply it by the number of emitting $e^{-}$ at $\gamma_{\rm m}$, $N_{\rm e, \, iso}$ (see \citealt{Sari_1998} and Eq.~\ref{eq: N_iso}). For a relativistic shock the number of emitting electrons in the emitting region is $N_{\rm e} = N_{\rm e, \, iso}/4\Gamma^2$ \citep{BDNP13}.}
\begin{align}
    F_{\rm \nu, \, max} \equiv P_{\rm \nu, \, max} N_{\rm e, \, iso} \frac{1}{4 \pi d_{\rm l}^2} \left( 1 + z \right) = \frac{m_{\rm e} c^2 \sigma_{\rm T}}{3 q_{\rm e}} \Gamma B N_{\rm e, \, iso} \frac{1}{4 \pi d_{\rm l}^2} \left( 1+z \right)
\end{align}
where $q_{\rm e}$ is its electric charge, $c$ is the speed of light, $N_{\rm e, \, iso}$ is the isotropic equivalent number of emitting electrons. We assume that the number of emitting electrons at $\gamma_{\rm m}$ can be parameterized as: 
\begin{align}
\label{eq: N_iso}
    N_{\rm e, \, iso} = N_{\rm 0} \times \frac{4 \pi}{3} f R^3
\end{align}
where $\frac{4 \pi}{3} f R^3$ is the emitting volume and $f$ is the volume filling factor. 

For different ordering of $\nu_{\rm m}, \nu_{\rm sa}, \nu_{\rm c}$, the flux density at $\nu_{\rm sa}$, $F_{\rm \nu_{\rm sa}}$, will be a different function of these frequencies, $F_{\rm \nu, \, max}$, and $p$ (based on the different spectral shapes discussed in \cite{granot_2002}):
\begin{align}
\label{eq: f_nu_cases}
    F_{\rm \nu_{\rm sa}} = \begin{cases}
F_{\rm \nu, \, max} \left( \frac{\nu_{\rm sa}}{\nu_{\rm m}}\right)^{1/3} & \nu_{\rm sa} < \nu_{\rm m} < \nu_{\rm c} \\
F_{\rm \nu, \, max} \left( \frac{\nu_{\rm sa}}{\nu_{\rm m}}\right)^{-(p-1)/2} & \nu_{\rm m} < \nu_{\rm sa} < \nu_{\rm c} \\
F_{\rm \nu, \, max} \left( \frac{\nu_{\rm c}}{\nu_{\rm m}}\right)^{-(p-1)/2} \left( \frac{\nu_{\rm sa}}{\nu_{\rm c}}\right)^{-p/2} & \nu_{\rm m}, \nu_{\rm c} < \nu_{\rm sa}
\end{cases}
\end{align}

Now we can derive the synchrotron self-absorption frequency by requiring that the flux at $\nu_{\rm sa}$ (defined in Eq.\ref{eq: f_nu_cases}) is equal to the blackbody flux given by:
\begin{align}
\label{eq: BB_flux}
    F_{\rm \nu, \, BB} & = \frac{2 \nu^2}{c^2} k_{\rm B} T f_{\rm A} \frac{R^2}{d_{\rm l}^2} \Gamma \left( 1+z\right)^3
\end{align}
where $k_{\rm B}$ is Boltzmann's constant, $T$ is the blackbody temperature, $f_{\rm A}$ is the area filling factor, and considering that at the low end of the energy distribution $k_{\rm B} T = \gamma m_{\rm e} c^2$. This procedure gives
\begin{align}
    \nu_{\rm sa} = \begin{cases}
\left[\bar{A} \frac{N_{\rm e, \, iso}B}{\gamma_{\rm m} f_{\rm A} R^2 \left( 1+z \right)^2 \nu_{\rm m}^{1/3}} \right]^{3/5} & \nu_{\rm sa} < \nu_{\rm m} < \nu_{\rm c} \\
\left[\bar{A} \bar{B} \frac{N_{\rm e, \, iso} B^{3/2} \Gamma^{1/2}}{f_{\rm A} R^2 \left( 1+z \right)^{5/2} \nu_{\rm m}^{-(p-1)/2}} \right]^{2/(p+4)} & \nu_{\rm m} < \nu_{\rm sa} < \nu_{\rm c} \\
\left[\bar{A} \bar{B} \frac{N_{\rm e, \, iso} B^{3/2} \Gamma^{1/2}}{f_{\rm A} R^2 \left( 1+z \right)^{5/2} \nu_{\rm m}^{-(p-1)/2} \nu_{\rm c}^{-1/2} } \right]^{2/(p+5)} & \nu_{\rm m}, \nu_{\rm c} < \nu_{\rm sa}
\end{cases}
\end{align}
where, for convenience, we defined $\bar{A} \equiv \frac{\sigma_{\rm T} c^2}{24 \pi q_{\rm e}}$ and $\bar{B} \equiv \left( \frac{2 \pi m_{\rm e} c}{q_{\rm e}} \right)^{-1/2}$. Finally, the observed peak of the SED (at the intersection between optically thin and thick regimes) is
\begin{align}
\label{eq: fp_nup_multi}
    F_{\rm p} = \begin{cases}
F_{\rm \nu, max} \, \, {\rm at} \, \, \nu_{\rm p} = \nu_{\rm m} & \nu_{\rm sa} < \nu_{\rm m} < \nu_{\rm c} \\
F_{\rm \nu, \, max} \left( \frac{\nu_{\rm sa}}{\nu_{\rm m}}\right)^{-(p-1)/2}  \, \, {\rm at} \, \, \nu_{\rm p} = \nu_{\rm sa} & \nu_{\rm m} < \nu_{\rm sa} < \nu_{\rm c} \\
F_{\rm \nu, \, max} \left( \frac{\nu_{\rm c}}{\nu_{\rm m}}\right)^{-(p-1)/2} \left( \frac{\nu_{\rm sa}}{\nu_{\rm c}}\right)^{-p/2}  \, \, {\rm at} \, \, \nu_{\rm p} = 
\nu_{\rm sa} & \nu_{\rm m}, \nu_{\rm c} < \nu_{\rm sa}
\end{cases}
\end{align}

\section{Off-axis equipartition analysis for electrons that radiate at the minimal frequency}
\label{sec: off_axis_derivation}

\cite{Matsumoto_2023} generalized the traditional equipartition analysis introduced in \cite{BDNP13} for a relativistic jetted emitter observed off-axis. However, their formalism, as mentioned in \cite{BDNP13} is only for electrons that radiate at a peak frequency $\nu_{\rm p}$. These electrons are likely to carry most of the relativistic electron energy only if $\nu_{\rm p} = \nu_{\rm m}$. However, if $\nu_{\rm m} < \nu_{\rm sa}$, most of the energy is carried by electrons with the minimal Lorentz factor, $\gamma_{\rm m}$, and the emission of which is self-absorbed, which means that we would be underestimating the energy without taking these electrons into account. We account for this additional energy by introducing a factor of $\left( \gamma_{\rm m}/\gamma_{\rm e} \right)^{2-p}$ to the energy of the electrons (as discussed in \S4.2.1 in \citealt{BDNP13}), and use it to generalize the process introduced in \cite{Matsumoto_2023} for any value of power-law index $p$.

Eq. (8), (9), and (10) in \cite{Matsumoto_2023} give
\begin{align}
    \label{eq: gamma_e_off}
    \gamma_{\rm e} = \frac{3 F_{\rm p} d_{\rm L}^2 \eta^{\frac{5}{3}}\Gamma^2}{2 \pi \nu_{\rm p}^2 \left( 1 + z \right)^3 m_{\rm e} f_{\rm A} R^2 \delta_{\rm D}} = 520 \left[ \frac{F_{\rm p, mJy} d_{\rm L, 28}^2 \eta^{\frac{5}{3}}}{\nu_{\rm p, 10}^2 \left( 1 + z \right)^3} \right] \frac{\Gamma^2}{f_{\rm A} R_{\rm 17}^2 \delta_{\rm D}}
\end{align}
\begin{align}
    \label{eq: Ne_off}
    N_{\rm e} = \frac{9 c F_{\rm p}^3 d_{\rm L}^6 \eta^{\frac{10}{3}}\Gamma^4}{2 \sqrt{3} \pi^2 q_{\rm e}^2 m_{\rm e}^2 \nu_{\rm p}^5 \left( 1 + z \right)^8 f_{\rm A}^2 R^4 \delta_{\rm D}^4} = 4.1 \times 10^{54} \left[ \frac{F_{\rm p, mJy}^3 d_{\rm L, 28}^6 \eta^{\frac{10}{3}}}{\nu_{\rm p, 10}^5 \left( 1 + z \right)^8} \right] \frac{\Gamma^4}{f_{\rm A}^2 R_{\rm 17}^4 \delta_{\rm D}^4}
\end{align}
\begin{align}
    \label{eq: bfield_off}
    B = \frac{8 \pi^3 m_{\rm e}^3 c \nu_{\rm p}^5 \left( 1 + z \right)^7 f_{\rm A}^2 R^4 \delta_{\rm D}}{9 q_{\rm e} F_{\rm p}^2 d_{\rm L,}^4 \eta^{\frac{10}{3}}\Gamma^4} = 1.3 \times 10^{-2} \, {\rm G} \left[ \frac{\nu_{\rm  p,10}^5 \left( 1 + z \right)^7}{F_{\rm p, mJy}^2 d_{\rm L,28}^4 \eta^{\frac{10}{3}}} \right] \frac{f_{\rm A}^2 R_{\rm 17}^4 \delta_{\rm D}}{\Gamma^4}
\end{align}
The energy in the electrons is
\begin{align}
    E_{\rm e} & = N_{\rm e} \gamma_{\rm e} m_{\rm e} c^2 \Gamma \left( \gamma_{\rm m}/\gamma_{\rm e} \right)^{2-p}
    \\
    \nonumber & = 3.36 \times 10^{48} \, {\rm erg} \times \left[ 520 \right]^{p-1} F_{\rm p, mJy}^{p+2} d_{\rm L,28}^{2\left( p+2 \right)} \nu_{\rm p}^{-\left( 2p+3 \right)} \left( 1 + z \right)^{-\left(3p+5\right)} \gamma_{\rm m}^{2-p} f_{\rm A}^{-\left( p + 1 \right)} R_{\rm 17}^{-\left(2p + 2 \right)} \Gamma^{2p+3} \delta_{\rm D}^{-\left( p + 3 \right)}
\end{align}
and in the magnetic field
\begin{align}
    E_{\rm B} & = \frac{\left( B \Gamma \right)^2}{8 \pi} V
    = 2.11 \times 10^{46} \, {\rm erg} \times F_{\rm p,mJy}^{-4} d_{\rm L, 28}^{-8} \nu_{\rm p, 10}^{10} \left( 1 + z \right)^{14} f_{\rm A}^{4} f_{\rm V} R_{\rm 17}^{11} \Gamma^{-10} \delta_{\rm D}^{2}
\end{align}
We next minimize the total energy with respect to the radius, and the energy is minimized at $E_{\rm B} = \left(\frac{2p+2}{11} \right) E_{\rm e}$. The equipartition radius is
\begin{align}
\label{eq: radius_off_derivation}
    R_{\rm eq} = \left( 10^{17} \, {\rm cm}\right) \left[ 159 \times\left[520\right]^{p-1} \right]^{\frac{1}{2p+13}} F_{\rm p, mJy}^{\frac{p+6}{2p+13}} d_{\rm L, 28}^{\frac{2\left( p+6 \right)}{2p+13}} \nu_{\rm p, 10}^{-1} \left( 1 + z \right)^{-\frac{3p + 19}{2p+13}} \epsilon^{\frac{1}{2p+13}} \gamma_{\rm m}^{\frac{2-p}{2p+13}} f_{\rm A}^{-\frac{p + 5}{2p+13}} f_{\rm V}^{-\frac{1}{2p+13}} \Gamma \delta_{\rm D}^{- \frac{p + 5}{2p+13}}
\end{align} 
and the equipartition energy
\begin{align}
\label{eq: energy_off_derivation}
    E_{\rm eq} = & \left( 3.36 \times 10^{48} \, {\rm erg}\right) \left[ 159 \right]^{-\frac{2 \left( p+1\right)}{2p+13}} \left[520\right]^{\frac{11\left(p-1\right)}{2p+13}} F_{\rm p, mJy}^{\frac{3p+14}{2p+13}} d_{\rm L, 28}^{\frac{2\left( 3p+14 \right)}{2p+13}} \nu_{\rm p, 10}^{-1} \left( 1 + z \right)^{-\frac{5p + 27}{2p+13}} \times
    \\
    \nonumber & \left[ 1 + \left( \frac{2p+2}{11} \right) \epsilon \right] \epsilon^{\frac{-2\left(p+1\right)}{2p+13}} \gamma_{\rm m}^{\frac{11 \left(2-p\right)}{2p+13}} f_{\rm A}^{-\frac{3\left(p + 1\right)}{2p+13}} f_{\rm V}^{\frac{2\left(p+1\right)}{2p+13}} \Gamma \delta_{\rm D}^{- \frac{7p + 29}{2p+13}}
\end{align}
Where $\epsilon = \left( \frac{\epsilon_{\rm B}}{\epsilon_{\rm e}} \right)/\left( \frac{2p+2}{11} \right)$, and $\epsilon_{\rm e}$ and $\epsilon_{\rm B}$ are introduced to account for deviations from equipartition (similar to \citealt{BDNP13}). We also introduce factors of $\xi^{\frac{1}{2p+13}}$ and $\xi^{\frac{11}{2p+13}}$, with $\xi = 1 + \epsilon_{\rm e}^{-1}$, to Eq. \ref{eq: radius_off_derivation} and \ref{eq: energy_off_derivation}, respectively, to account for hot protons (see discussion in \S4.2.2 in \citealt{BDNP13}). Finally, it is important to note that these definitions of $\epsilon_{\rm B}$ and $\epsilon_{\rm e}$ are not as fraction of the post-shock energy and therefore are different from the definitions introduced for the non-relativistic case (Appendix \S\ref{sec: ssa_frequency} and Eq. \ref{eq: density_bfield}).

\bibliography{sample631}{}
\bibliographystyle{aasjournal}

\end{document}